\documentclass[nofootinbib,aps,twocolumn,superscriptaddress,showpacs,english]{revtex4-1}

\usepackage[T1]{fontenc}
\usepackage[utf8]{inputenc}

\usepackage[english]{babel}
\usepackage{amsmath}
\usepackage{amssymb}
\usepackage{wasysym}
\usepackage{graphicx}
\usepackage{xcolor}
\usepackage{bm} 
\usepackage{cancel}
\usepackage{mdframed}
\usepackage{soul}
\usepackage{todonotes}

\usepackage{amsmath}
\usepackage{tikz}
\usetikzlibrary{arrows,matrix,positioning}
\usepackage{comment}

\usepackage{braket}

\usepackage[linktocpage=true,
  colorlinks=true, 
  pdfborder={0 0 0},
  linkcolor=blue,
  citecolor=red,
  filecolor=yellow,
  urlcolor=blue,
  bookmarks,
  pdfauthor={},
]{hyperref}

\usepackage{xcolor}   
\usepackage{pst-node} 

\usepackage{pstricks-add}

\usepackage{ulem}
\definecolor{gray}{RGB}{100,100,100}

\newcommand{\DAG}[1]{#1^{\dagger}}

\newcommand{\tikzmark}[2]{\tikz[overlay,remember picture] \node[minimum width=1.5em] (#1) {#2};}

\newcommand{\rev}[1]{{#1}} 

\newcommand{\im}{{\mathrm{i}}}

\newcommand{\opa}{{\mathop{\hat{a}}}}
\newcommand{\opc}{{\mathop{\hat{c}}}}
\newcommand{\opad}[1]{{{\mathop{\hat{a}_{#1}^\dagger}}}} %

\newcommand{\opn}{{\mathop{\hat{n}}}}

\newcommand{\opH}{{\mathop{\hat{H}}}}

\newcommand{\mybf}[1]{#1}

\newcommand{\mybfone}{{\mybf 1}} 
\newcommand{\mybfW}{{\mybf W}} 
\newcommand{\mybfS}{{\mybf S}} 
\newcommand{\mybfU}{{\mybf U}}

\newcommand{\mybfV}{{\mybf V}}

\newcommand{\mybfH}{{\mybf H}}

\newcommand{\mybfG}{{\mybf G}}
\newcommand{\mybfX}{{\mybf X}}
\newcommand{\mybfSigma}{{\mybf \Sigma}}

\newcommand{\mybfDelta}{{\mybf \Delta}}

\newcommand{\Tr}{\text{~Tr}}

\newcommand{\GrazTh}{Institute of Theoretical and Computational Physics, Graz University of Technology, NAWI Graz, 8010 Graz, Austria}

\begin{document}

\title[First-principles quantum transport simulation of CuPc on Au(111) and Ag(111)]{First-principles quantum transport simulation of CuPc on Au(111) and Ag(111)}

\renewcommand{\thefootnote}{\fnsymbol{footnote}}

\author{M. Rumetshofer}  
\email{m.rumetshofer@tugraz.at} 
\affiliation{\GrazTh} 
\author{D. Bauernfeind}    
\affiliation{\GrazTh}
\author{E. Arrigoni}    
\affiliation{\GrazTh}
\author{W. von der Linden}    
\affiliation{\GrazTh}

\date{\today}

\begin{abstract}
We investigate equilibrium and transport properties of a copper phthalocyanine (CuPc) molecule adsorbed on Au(111) and Ag(111) surfaces.
The CuPc molecule has essentially three localized orbitals close to the Fermi energy resulting in strong local Coulomb repulsion not accounted for properly in density functional calculations. Hence, they require a proper many-body treatment within, e.g., the Anderson impurity model (AIM).
The occupancy of these orbitals varies with the substrate on which CuPc is adsorbed.
Starting from density functional theory calculations, we determine the parameters for the AIM embedded in a noninteracting environment that describes the residual orbitals of the entire system.
While correlation effects in CuPc on Au(111) are already properly described by a single orbital AIM, for CuPc on Ag(111) the three orbital AIM problem can be simplified into a two orbital problem coupled to the localized spin of the third orbital. This results in a Kondo effect with a mixed character, displaying a symmetry between SU(2) and SU(4). The computed Kondo temperature is in good agreement with experimental values. To solve the impurity problem we use the recently developed fork tensor product state solver. 
To obtain transport properties, a scanning tunneling microscope (STM) tip is added to the CuPc molecule absorbed on the surface.
{We find that the  transmission depends on the detailed position of the STM tip above the CuPc molecule in good agreement with differential conductance measurements.}
\end{abstract}

\maketitle

\section{Introduction}
\begin{figure}
\begin{flushleft}
 \hspace{0\columnwidth}(a) \hspace{0.5\columnwidth}  (b) 
\end{flushleft}
 \begin{center}
 \includegraphics[width=0.45\columnwidth,angle=0]{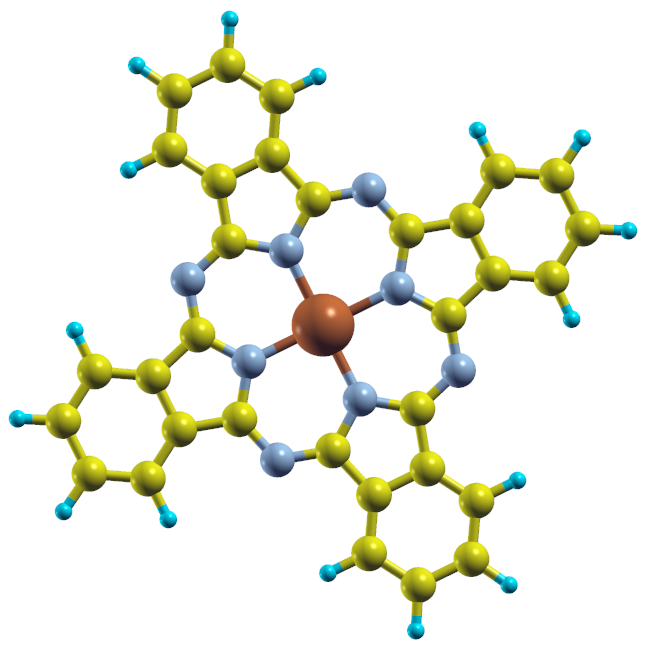} 
 \includegraphics[width=0.45\columnwidth,angle=0]{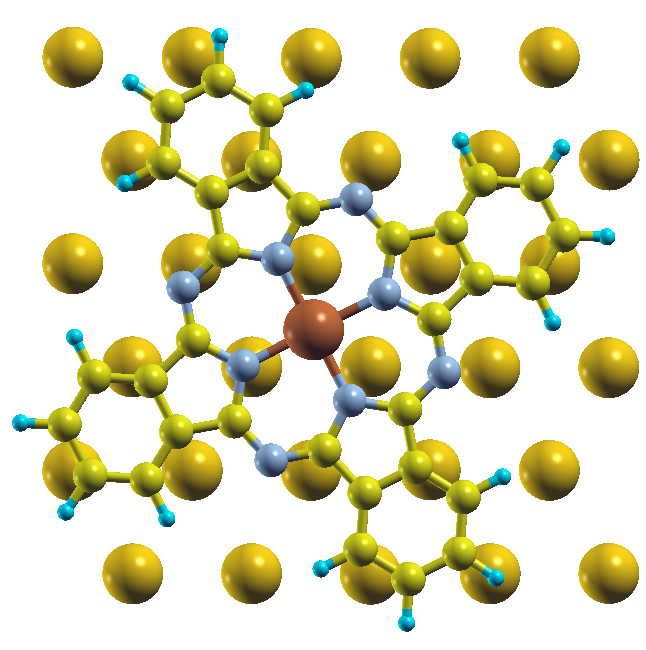}
 \caption{An isolated CuPc molecule (a) and the position of the CuPc molecule on the Au(111) surface (b). The pictures are drawn with XCrySDen \cite{Kokalj_xcrysden_2003}. Color code: C atoms, yellow; H atoms, cyan; N atoms, gray; Cu atom, red; Au atoms, gold.}
\label{fig:CuPc}
\end{center}
\end{figure}

Copper phthalocyanines (CuPc) 
$
\text{C}_{32}\text{H}_{16}\text{CuN}_{8}
$
are magnetic, organic, semiconducting molecules with a brilliant blue color~\cite{Brink_MagneticBlue_2007,Schwieger_SemoCondCuPc_2002}.
An isolated CuPc molecule is depicted in Fig. \ref{fig:CuPc}(a).
The electronic properties of transition metal phthalocyanines (TMPc) in general have been studied extensively in several environments using different methods, in both, experiment and theory, e.g.: pristine CuPc \cite{Aristov_pristineCuPc_2007,Marom_CuPcwithGW_2011}, TMPc between monoatomic chains \cite{Nazin_CuPcAuchain_2003,Calzolari_MPc_2007,Fadlallah_FlourMPc_2016}, or TMPc on metal surfaces \rev{\cite{Zhenpeng_MPcAu111_2008,Stepanow_CuPc_2010,Stepanow_CuPc_2011,Franke_MPcPb111_2011,Soriano_MPcPb111_2012,Ziroff_Phthalocyanines_2012,Mugarza_CuPc_2012,Salomon_CoPc_2013,Jacob_MPcPb111_2013,Wruss_Phthalocyanines_2014}}.
\rev{Especially in the latter, one of the cooperative many-body phenomena in solid state physics, the Kondo effect \cite{Kondo_1964,Hewson_Kondo_1997,Goldhaber-Gordon_SET_1998,Park_Kondo_2002}, has been observed \cite{Mugarza_CuPc_2011,Korytar_CuPcKondo_2011,Minamitani_FePcKondo_2012,Hiraoka_exptheoFeKondo_2017}.}

In the {gas phase}, the transition metal (TM) in TMPc binds to four isoindole ligands leaving the ion in a [TM]$^{2+}$ state.
The molecule itself has a square planar D$_{4h}$ symmetry and hence the TM
$d$ states transform as $b_{2g}~(d_{xy})$, $b_{1g}~(d_{x^2-y^2})$, $a_{1g}~(d_{z^2})$, and $e_{g}~(d_{xz}, d_{yz})$.
Depending on their symmetry and energetic position, these orbitals \rev{hybridize} to a different degree with $p$ orbitals of the C and N atoms.
In the gas phase CuPc has a total spin $S = 1/2$ due to one unpaired electron in the $b_{1g}$ state. The highest occupied and lowest unoccupied molecular orbital (HOMO and LUMO) are delocalized $a_{1u}$ and $2e_{g}$ {$\pi$ orbitals} with marginal contributions from the TM $d$ states and therefore mainly located at the Pc.
If the molecule is adsorbed on Ag(100) \cite{Mugarza_CuPc_2011}, the surface charge transfer from the metal surface to the $2e_g$ states generates another unpaired spin $S = 1/2$ at the Pc.
Therefore, in the adsorbed molecule one finds two weakly interacting spins, one localized on the Cu orbitals ($b_{1g}$ state) and the other induced in the Pc ($2e_g$ states), leading to singlet ($S=0$) and triplet ($S=1$) states of the molecule.
Such a charge transfer between the surface and the $2e_g$ states does not occur in CuPc on Au(111) \cite{Wruss_Phthalocyanines_2014}, where the molecule remains in the doublet ($S = 1/2$) state. 

Photoelectron spectroscopy (PES) measurements~\cite{Ziroff_Phthalocyanines_2012,Wruss_Phthalocyanines_2014} for CuPc on Au(111) and Ag(111) show a sharp structure at the Fermi energy for CuPc on Ag(111) but not for CuPc on Au(111).
A generalized Kondo scenario in the $2e_g$ states is suggested to be the possible origin.
Mugarza {\it et al.} \cite{Mugarza_CuPc_2011,Mugarza_CuPc_2012} measured the differential conductance of CuPc on Ag(100) at different tip positions of the scanning tunneling microscope (STM) and found a Kondo resonance 
in the $2e_g$ orbitals and estimated the Kondo temperature to $T_K=27\pm2~$K.
Korytár {\it et al.} \cite{Korytar_CuPcKondo_2011} performed density functional theory (DFT) calculations for CuPc on Ag(100) using localized Wannier functions, and employing the noncrossing approximation (NCA) to solve the multiorbital Anderson impurity model (AIM) \cite{Anderson_AIM_1961} describing the $2e_g$ states plus exchange interaction with the single occupied $b_{1g}$ state. 
Korytár {\it et al.} \cite{Korytar_CuPcKondo_2011} were not able to estimate the Kondo temperature from their {\it ab initio} calculations and state that the underlying reason for this is the DFT level misalignment due to the lack of Coulomb repulsion.
Here, we will present evidence that the \rev{hybridization} strength is a much more important reason for the discrepancy of the Kondo temperature found in experiment and theory.
It is well known that the  Kondo temperature depends sensitively (exponentially) on the \rev{hybridization} strength with the environment. Since the latter depends on the adsorption geometry of the molecule on the respective metal surface, reliable estimates for the Kondo temperature can only be found if the correct geometry for the underlying DFT calculation is used.

\rev{In this paper we calculate the transport properties of CuPc on Au(111) and Ag(111) from first principles to determine a simplified model, sufficient for the description of the system.
This is important in order to predict situations where the Kondo effect can be observed as well as its properties: Kondo temperature, symmetry, involved orbitals.
Especially possibilities for the experimental observation of the Kondo cloud are a longstanding question \cite{Affleck_KondoCloud_2001,Fu_KondoCloud_2007,Affleck_FriedelKondoCloud_2008,Park_KondoCloud_2013} where {\it ab initio} calculations can help gaining deeper understanding.
Contrary to Korytár {\it et al.} \cite{Korytar_CuPcKondo_2011}, for our calculations} we use the optimized adsorption geometry obtained by Huang {\it et al.} \cite{Wruss_Phthalocyanines_2014} and check that our (many-body) spectral functions are consistent with the density of states (DOS) obtained by Heyd-Scuseria-Ernzerhof (HSE) DFT calculations and by ultraviolet photoemission spectroscopy (UPS) experiments performed in Ref. \cite{Wruss_Phthalocyanines_2014}.
This allows us to estimate the Kondo temperature from first principles and to calculate the qualitative behavior of the differential conductance in the STM measurements \cite{Mugarza_CuPc_2011,Mugarza_CuPc_2012}.
For the inclusion of many-body effects, we apply the method suggested by Droghetti {\it et al.} \cite{Droghetti_TOV_2017} to construct an effective  Anderson impurity model (AIM).
For CuPc on Au(111) this yields a single orbital AIM for the copper $b_{1g}$ orbital.
In the case of CuPc on Ag(111), we obtain a three orbital AIM for the copper $b_{1g}$ and the two nearly degenerate $2e_g$ orbitals mainly located at the Pc.
Due to the negligible \rev{hybridization} of the $b_{1g}$ orbital with the remaining orbitals it can be treated in the atomic limit. The resulting exchange coupling to the $2e_g$ orbitals can then be accounted for in mean field approximation. Eventually, this leads to an effective spin-dependent energy shift for the  electrons in  the $2e_g$ states. 
For the many-body treatment of the physics in the degenerate $2e_g$ orbitals we use the recently developed fork tensor product state (FTPS) solver \cite{Bauernfeind_FTPS_2017,Bauernfeind_FTPSSMO_2018,Bauernfeind_THESIS_2018}. 
Our calculations yield  a reliable {\it ab initio} estimate of the Kondo temperature and reproduce the qualitative behavior of the differential conductance found in the STM measurements in Refs. \cite{Mugarza_CuPc_2011,Mugarza_CuPc_2012}.
For the computation of the coherent contributions to the equilibrium transmission, we treat the leads in the wide-band limit, to suppress lead-induced effects.

This paper is structured as follows. Section \ref{sec:method} introduces the methods employed, i.e., the DFT calculation, the mapping onto and solution of the AIM, as well as transport calculations. In Sec.~\ref{sec:results} we apply this approach to CuPc on Au(111) and Ag(111) and discuss the results. 

\section{Method and computational details}
\label{sec:method}

We perform DFT calculations for CuPc on Au(111) and Ag(111), respectively, to obtain the one particle Hamiltonian of these systems. 
Technical details of the DFT calculations are presented in Sec. \ref{sec:DFT}.
For a reliable description of the system it is essential to include correlation effects of the molecule.  
To this end we construct an appropriate AIM based on the DFT orbitals, see Sec. \ref{sec:AIM}. We calculate the self-energy of this AIM employing the FTPS solver, which is briefly discussed in Sec.~\ref{sec:solver}. Finally, in Sec.~\ref{sec:trans} we discuss how to combine the one particle with the many-body part in transport calculations.


\subsection{Density functional calculations}
\label{sec:DFT}

\begin{figure} 
 \begin{center}
 \includegraphics[width=1.00\columnwidth,angle=0]{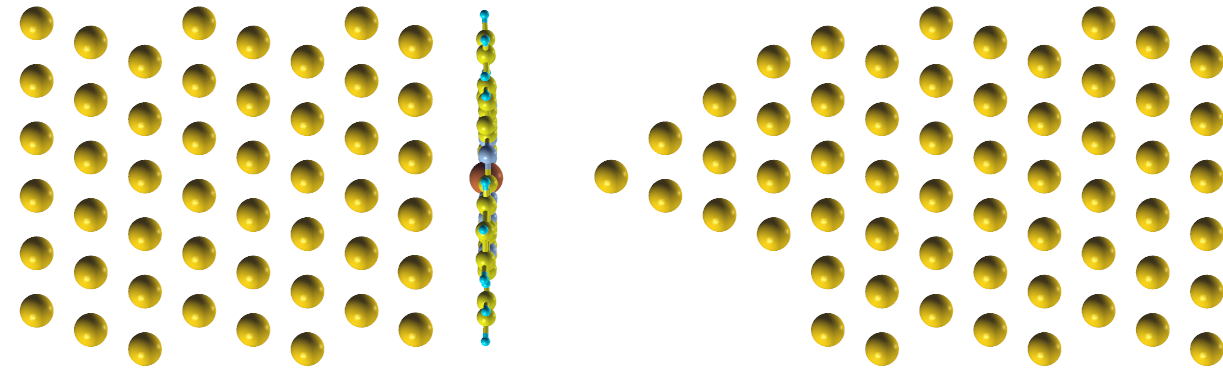} 
 \end{center}
 \caption{Two-dimensional cut through the central region for the simulation of the CuPc molecule sandwiched between an  Au(111) substrate and the STM tip. The picture is drawn with XCrySDen \cite{Kokalj_xcrysden_2003}.}
\label{fig:CuPc-AuSTM}
\end{figure}

To  determine the one particle part of the Hamiltonian, we perform DFT calculations for CuPc on Au(111) and Ag(111), respectively, in STM configuration. 
In such an STM configuration the CuPc molecule is sandwiched between the (111) surface of the Au/Ag substrate and an STM tip, see Fig. \ref{fig:CuPc-AuSTM}. The molecule lies in the $xy$ plane, which is defined by the surface of the substrate. The $z$ axis, perpendicular to the surface, defines the transport direction.
To model the tip, we use a tetrahedron attached to a three-dimensional semi-infinite system, both of the same material.

For the transport calculations, the system is split into a central region
and two {\it leads}. The central region, displayed in Fig.~\ref{fig:CuPc-AuSTM}, consists of the CuPc molecule, the actual tip, and eight layers of the substrate material on each side. On both sides, this central region is attached to the residual parts of the semi-infinite systems, which we will denote as {\it leads} (not shown in Fig.~\ref{fig:CuPc-AuSTM}).

For the DFT calculation, it is necessary to have a periodic system
in the $xy$ plane, which is therefore split into appropriate unit cells.
According to \cite{Wruss_Phthalocyanines_2014} we use a lattice constant of $4.18$ \AA~for Au and $4.15$ \AA~for Ag and p($6\times 5$) Au(111) and Ag(111) surfaces.
We chose the tip material to be the same as the surface, i.e., an Ag tip for the Ag(111) surface and an Au tip for the Au(111) surface. We want to emphasize though that one could also use any other tip material.
To reduce the influence of the tip onto the molecule, we choose the molecule-tip distance to be large ($5.57~$\AA~in the Au and $5.89~$\AA~in the Ag setup) compared to the distance between molecule and surface.

The relaxation of molecules on surfaces is generally a highly nontrivial task and, moreover, the molecular position strongly influences electrical, magnetic and transport properties.
Therefore, we use the optimized adsorption geometries from Huang {\it et al.} \cite{Wruss_Phthalocyanines_2014}. The resulting positions of the CuPc molecule on the Au(111) and Ag(111) surfaces are shown in Figs. \ref{fig:CuPc}(b) and \ref{fig:tippos}, respectively.
\rev{Importantly, the distance between the molecule and the relaxed surface layer of the Ag(111) surface is $2.84~$\AA, which is larger than the distance obtained in Refs. \cite{Mugarza_geom_2010,Korytar_CuPcKondo_2011}. In Sec. \ref{sec:Kondo} we discuss the influence of this discrepancy on the estimation of the Kondo temperature.}

\rev{The DFT calculations are performed with {\sc SIESTA} \cite{Soler_Siesta_2002} and Tran{\sc SIESTA} \cite{Brandbyge_TranSiesta_2002} using the Perdew-Burke-Ernzerhof (PBE) \cite{Perdew_BPE_1996} functional.
We exclusively perform spin-unpolarized DFT calculations, since we want to describe the magnetic properties using an additional strongly correlated many-body Hamiltonian.
Calculation details are given in Appendix \ref{sec:App:DFT}.}

\subsection{Projection onto the AIM}
\label{sec:AIM}
In this section we present the formalism to construct an AIM in orthogonal orbitals starting from the DFT results. The AIM then allows us to study correlation effects on the quantum transport in addition to those covered already by the DFT.
We will give a concise introduction to the approach proposed by Droghetti {\it et al.} \cite{Droghetti_TOV_2017}, along with some modifications.

\begin{figure}
\begin{center}
 \vspace{0.5cm}
 \includegraphics[width=1.00\columnwidth,angle=0]{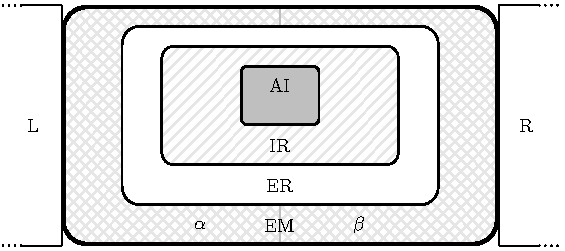} 
 \end{center}
  \caption{The schematic representation of the transport region consisting of left (L) and right (R) lead and the extended molecule (EM), or rather
the central region, with its subsystems: the extended region (ER), the interacting region (IR), and the Anderson impurity (AI).
}
\label{fig:schema}
\end{figure}

In a first step, the system is separated into a noninteracting (coherent) part and a strongly correlated part described by the AIM.
Therefore, Droghetti {\it et al.}~\cite{Droghetti_TOV_2017} divide the system into several regions (see Fig.~\ref{fig:schema}).
The left lead (L) couples to the so-called extended molecule (EM), which in turn is coupled to the right lead (R). The extended molecule is often also referred to as the central region. The leads are chosen such that there is no single particle overlap between them. In our case, we choose the left lead being on the metal surface side and the right lead on the tip side of the system. The EM is further subdivided into: extended region (ER), interacting region (IR) and the Anderson impurity (AI), with $\text{EM}\supseteq\text{ER}\supseteq\text{IR}\supseteq\text{AI}$.
{The IR includes all orbitals that may contribute to the AI.
From the IR, we determine the AI by diagonalizing $\mybfH_{\text{IR}}$ and selecting the correlated orbitals depending on their localization and filling.} Hence, the AI describes the strongly correlated orbitals for which a Hubbard interaction is taken into account.
In addition to the CuPc molecule we include orbitals from the actual tip, as well as the first surface layer of the substrate as IR.
This allows for the possibility that the orthogonal correlated orbitals extend into the tip or the surface.
In the following we have to construct an AI with basis functions orthonormal to the rest of the system. 
Therefore, we define the ER as consisting of all orbitals with finite overlap with the IR, including the IR itself. 
The set of remaining orbitals (which we denote as $\text{EM}\setminus\text{ER}$) are split into two parts, that couple to the left ($\alpha$) or the right ($\beta$) lead, respectively. 

Since {\sc SIESTA} uses atomic orbitals, we have to take their nonorthogonality into account.
Green’s function theory for nonorthogonal basis functions is for example discussed in Ref. \cite{Thygesen_nonorthobasis_2006}.
The overlap matrix $\mybfS$ and the single particle Hamiltonian $\mybfH$ of the EM have the structure
\begin{align}
 \mybfX = \begin{pmatrix}\mybfX_{\alpha\alpha} & \mybfX_{\alpha\text{ER}} & \mybfX_{\alpha\beta} \\
            \mybfX_{\alpha\text{ER}}^\dagger & \mybfX_{\text{ER}} & \mybfX_{\beta\text{ER}}^\dagger \\
            \mybfX_{\alpha\beta}^\dagger & \mybfX_{\beta\text{ER}} & \mybfX_{\beta\beta}\end{pmatrix} \;,
            \label{eq:X}
\end{align}
with $X$ denoting either $S$ or $H$.
{$\mybfX_{ij}^\dagger$ is the conjugate transpose of the block matrix $\mybfX_{ij}$.}
To project onto an AI that is orthogonal to all remaining orbitals, we have to find a transformation $\mybfW$ that divides the ER into a noninteracting (NI) and an AI part such that:
\begin{align}
\label{eq:barS}
\bar\mybfS = \mybfW^\dag S \mybfW = \begin{pmatrix} \mybfone_{N_{\text{AI}}} & 0 & 0 & 0 \\
            0 & \tikzmark{a}{$\mybfS_{\alpha\alpha}$} & \bar\mybfS_{\alpha\text{NI}} & \mybfS_{\alpha\beta} \\
            0 & \bar\mybfS_{\alpha\text{NI}}^\dagger & \bar\mybfS_{\text{NI}} & \bar\mybfS_{\beta\text{NI}}^\dagger \\
            0 & \mybfS_{\alpha\beta}^\dagger & \bar\mybfS_{\beta\text{NI}} & \tikzmark{b}{$\mybfS_{\beta\beta}$}\end{pmatrix}
\end{align}
and
\begin{align}
\label{eq:barH}
 \bar\mybfH = \mybfW^\dag H \mybfW = \begin{pmatrix}  \epsilon_{\text{AI,D}} & 0 & \bar\mybfH_{\text{AI},\text{NI}} & 0 \\
            0 & \tikzmark{a}{$\mybfH_{\alpha\alpha}$} & \bar\mybfH_{\alpha\text{NI}} & \mybfH_{\alpha\beta}  \\
            \bar\mybfH_{\text{AI},\text{NI}}^\dagger & \bar\mybfH_{\alpha\text{NI}}^\dagger & \bar\mybfH_{\text{NI}} & \bar\mybfH_{\beta\text{NI}}^\dagger \\
            0 & \mybfH_{\alpha\beta}^\dagger & \bar\mybfH_{\beta\text{NI}} & \tikzmark{b}{$\mybfH_{\beta\beta}$}\end{pmatrix}.
\end{align}
The block in the upper left corner describes the AI in basis functions orthogonal to each other and to the noninteracting (NI) orbitals describing the rest of the ER.
Note that Hamiltonian and overlap matrix of $\text{EM}\setminus\text{ER}$ are unaffected by the transformation $W$, hence $\bar\mybfX_{ij} = \mybfX_{ij}$ for $i,j\in\{\alpha,\beta\}$.

$\mybfW$ is neither unitary nor uniquely defined. One possibility to obtain it is the following procedure using three consecutive transformations $\mybfW_1$, $\mybfW_2$, and $\mybfW_3$  with $\mybfW = \mybfW_1 \mybfW_2 \mybfW_3$:
\begin{align}
  \mybfW = \begin{pmatrix}0 & \mybfone_{N_\alpha} & 0 & 0 \\
            \mybfW_{\text{AI}} & 0 & \mybfW_{\text{NI}} & 0 \\
            0 & 0 & 0  & \mybfone_{N_\beta}\end{pmatrix}\;. \label{eq:TransformationW}
\end{align}
In the first step, the AI is projected out using:
\begin{align}
 \mybfW_1 = \begin{pmatrix}0 & \mybfone_{N_\alpha} & 0 & 0 \\
            \mybfU_{\text{ER}} & 0 & \mybfU_{\text{NI}} & 0 \\
            0 & 0 & 0  & \mybfone_{N_\beta}\end{pmatrix},
\end{align}
where $\mybfU_{\text{ER}}$ consists of the contributions of the orbitals in IR to the impurity orbitals. $\mybfU_{\text{NI}}$ is the identity matrix with removed columns at the indices of the impurity orbitals.
The second step orthogonalizes the AI to all other orbitals in ER by changing the orbitals in NI.
\begin{align}
 \mybfW_2 = \begin{pmatrix} \mybfone_{N_{\text{AI}}} & 0 & -\mybfW_{\text{SB}} & 0 \\
            0 & \mybfone_{N_\alpha} & 0 & 0 \\
            0 & 0 & \mybfone_{N_{\text{NI}}} & 0 \\
            0 & 0 & 0  & \mybfone_{N_\beta}\end{pmatrix}
            \label{eq:W2}
\end{align}
with $\mybfW_{\text{SB}} = \tilde \mybfS_{\text{AI}}^{-1}  \tilde \mybfS_{\text{AI,NI}}$, $\tilde \mybfS_{\text{AI}} = \mybfU_{\text{ER}}^\dagger \mybfS_{\text{ER}} \mybfU_{\text{ER}}$, and $\tilde \mybfS_{\text{AI,NI}} = \mybfU_{\text{ER}}^\dagger \mybfS_{\text{ER}} \mybfU_{\text{NI}}$.
To be precise, $\tilde\mybfX =\{\tilde\mybfS, \tilde\mybfH\}$ are the matrices after the first transformation step, i.e., $\tilde\mybfX = \mybfW_1^\dagger \mybfX \mybfW_1$,
and $\tilde \mybfS_{\text{AI}}$, $\tilde \mybfS_{\text{AI,NI}}$ and $\tilde\mybfH_{\text{AI}}$ are block matrices of $\tilde\mybfS$ and $\tilde\mybfH$, respectively.
The third step diagonalizes $\tilde\mybfS_{\text{AI}}$, and $\tilde\mybfH_{\text{AI}}$.
The AI basis functions are orthogonalized via a Löwdin transformation and $\tilde\mybfH_{\text{AI}}$ is diagonalized by solving the eigenvalue problem
\begin{align}
 \tilde\mybfS_{\text{AI}}^{-1/2} \tilde \mybfH_{\text{AI}} \tilde\mybfS_{\text{AI}}^{-1/2}\mybfU_{\text{AI},\psi} = \mybfU_{\text{AI},\psi} \epsilon_{\text{AI,D}}\;.
\end{align}
This results in the transformation
\begin{align}
 \mybfW_3 = \begin{pmatrix} \mybfW_{3,{\text{AI}}} & 0 & 0 & 0 \\
            0 & \mybfone_{N_\alpha} & 0 & 0 \\
            0 & 0 & \mybfone_{\text{NI}} & 0 \\
            0 & 0 & 0  & \mybfone_{N_\beta}\end{pmatrix}
\end{align}
with $ \mybfW_{3,{\text{AI}}} = \tilde\mybfS_{\text{AI}}^{-1/2}\mybfU_{\text{AI},\psi}$. 

At this point, we obtained a description of the EM with a strongly correlated AI only coupled to a subset of all orbitals. The only missing ingredient for transport calculations is the treatment of the leads L and R. As these leads are assumed to be noninteracting, they can be accounted for using \rev{hybridization} functions (also-called contact or tunneling self-energy).
Given the lead Hamiltonian $H_i$ with $i\in\{\text{L, R}\}$ and its coupling to the EM, $V_{\text{EM},i}(z) = \left( \mybfH_{\text{EM},i} - z\mybfS_{\text{EM},i}\right)$, the \rev{hybridization} function is defined by
\begin{align}
    \mybfDelta^{i}(z) = V_{\text{EM},i}(z) \frac{1}{z\mybfS_i - \mybfH_i} V_{i,\text{EM}}(z)\;.
    \label{eq:hyb}
\end{align}
{The retarded $\mybfDelta^{\text{r},i}(\omega)$ and advanced $\mybfDelta^{\text{a},i}(\omega)$ \rev{hybridization}s are obtained by replacing $z\rightarrow\omega\pm\im 0^+$} in Eq.~\ref{eq:hyb}.
With this, the (noninteracting) retarded Green's function projected in the EM subspace is given by
\begin{align}
    \mybfG^{\text{r},0}(\omega) = \frac{1}{(\omega+\im 0^+)\mybfS - \mybfH - \mybfDelta^{\text{r},\text{L}}(\omega) - \mybfDelta^{\text{r},\text{R}}(\omega)} \;.
    \label{eq:G}
\end{align}
{Note that the advanced Green's function can be obtained by $\mybfG^{\text{a},0}(\omega)= \left(\mybfG^{\text{r},0}\right) ^\dagger (\omega)$. We refer to} Ref. \cite{Ryndyk_Green_2009} for an introduction into Green's function techniques.
{In the following, we only discuss retarded quantities and, therefore, we neglect the superscripts for retarded and advanced and reintroduce them when required.}
As the Green's functions are the inverse of the noninteracting Hamiltonian  $H$ in a single particle basis, they have the same block-matrix structure, given by either Eq.~\ref{eq:X} in the original space or Eq.~\ref{eq:barH} in the transformed space.
Under transformations W of the Hamiltonian, Green's functions transform with the inverse of $\mybfW$ and therefore $\bar\mybfG(\omega) = \mybfW^{-1} \mybfG(\omega)  {\mybfW^{-1}}^\dagger$.
\rev{Hybridization}s on the other hand transform like $H$, i.e., according to $\bar\mybfDelta = \mybfW^\dagger \mybfDelta \mybfW$.
Especially, the Green's function of the extended region (ER) transforms like
\begin{align}
\mybfG_{\text{ER}} = \begin{pmatrix} \mybfW^{\dagger}_{\text{AI}} , \mybfW^{\dagger}_{\text{NI}} \end{pmatrix} \bar\mybfG_{\text{ER}} \begin{pmatrix} \mybfW_{\text{AI}} \\ \mybfW_{\text{NI}} \end{pmatrix}\;.
\end{align}
For our analysis, we need the DOS projected on orbital $\nu$ of the nonorthonormal atomic basis of the ER obtained from SIESTA.
\begin{align}
A_{\text{ER},\nu}(\omega) &= -\frac{1}{\pi} \Im \left( \mybfG_{\text{ER}}(\omega)\mybfS_{\text{ER}} \right) _{\nu\nu} 
\end{align}
The atomic-element resolved DOS is obtained by summation over all basis functions belonging to the corresponding atom. Similarly, the total DOS is the sum over all projections, $A_{\text{ER}}(\omega) = \sum_\nu A_{\text{ER},\nu}(\omega)$. 

Next, let us look at the strongly correlated part, i.e.: the AIM in more detail. 
The Hamiltonian of the isolated impurity is
\begin{align}
 \opH_\text{AI} = \sum_{i\sigma} \left(\epsilon_{\text{AI,D},i\sigma} - H^{\text{dc}}_{i\sigma} \right) \opn_{i\sigma} + \opH_\text{int},
\label{eq:modelH}
\end{align}
where 
\begin{align}
 \opH_\text{int} = \frac{1}{2}\sum_{ij\sigma} U_{ij}\opn_{i\sigma}\opn_{j\bar\sigma} + \frac{1}{2}\sum_{i\neq j,\sigma} V_{ij}\opn_{i\sigma}\opn_{j\sigma} \;.
\end{align}
Above, $\opn_{i\sigma} = \opa_{i\sigma}^\dagger \opa_{i\sigma}$ is the particle number operator of orbital $i$ and spin $\sigma$ in second quantization with creation (annihilation) operators $\opa_{i\sigma}^\dagger$ ($\opa_{i\sigma}$).
{We also assumed that non-density-density-terms are negligible.}
{In the  AIM, the impurity is coupled to a bath of noninteracting fermions:}
\begin{align}
    \opH_\text{AIM} = \opH_\text{AI} + \sum_{ik\sigma} \tilde V_{ik }\left( \opa_{i\sigma}^\dagger \opc_{ik\sigma} + h.c. \right) + \sum_{ik\sigma} \epsilon_{ik} \opn_{ik\sigma}\;.
    \label{eq:H_AIM}
\end{align}
$\opc_{ik\sigma}^\dagger$ ($\opc_{ik\sigma}$) are the creation (annihilation) operators of the $k$th bath state of orbital $i$ with spin $\sigma$. 
We have already determined the on-site energies via the transformation scheme Eqs. \ref{eq:barH} and \ref{eq:TransformationW}. For the double counting, we used the around mean field (AMF) double counting \cite{Karolak_arxiv_2010},
\begin{align}
H^{\text{dc}}_{i\sigma} &= x_i \sum_j  U_{{i}{j}}n_{{j}}^0 + x_i \sum_{{{j}}\neq {i}}  V_{{i}{{j}}}n_{{j}}^0\;,
\label{eq:dc}
\end{align}
where we introduced an orbital dependent factor $x_i$ according to Ref.~\cite{Haule_dc_2015}.
$n_{{j}}^0$ is the occupation of orbital $j$ obtained from DFT.
As the bath of free fermions is supposed to describe the NI, its \rev{hybridization} function $\bar\mybfDelta_{\text{AI}}(\omega)$ defines the bath parameters $\epsilon_{ik}$ and $\tilde V_{ik}$ via
\begin{align}\label{eq:bathParameters}
 \left\{ \bar\mybfDelta_{\text{AI}} (\omega) \right\}_{ii} &= \left\{ \bar\mybfH_{\text{AI,NI}}\bar\mybfG^{0}_{\text{NI}}(\omega)\bar\mybfH_{\text{NI,AI}} \right\}_{ii} \notag \\
 &\stackrel{!}{=} \sum_k \frac{ \tilde V_{ik}^2}{\omega+\im 0^+ - \epsilon_{ik}}\;,
\end{align}
i.e., we neglect off-diagonal \rev{hybridization}s $\left\{ \bar\mybfDelta_{\text{AI}} (\omega) \right\}_{ij}$ for $i\neq j$.
To clarify, for any given \rev{hybridization} function $\bar\mybfDelta_{\text{AI}}(\omega)$ we have to find bath parameters such that Eq.~\ref{eq:bathParameters} is satisfied.\footnote{\rev{We obtain the actual values of $\epsilon_{ik}$ and $\tilde V_{ik}$ by the following procedure. Starting from
the hybridization $\left\{ \bar\mybfDelta_{\text{AI}} (\omega) \right\}_{ii}$, we define equally spaced energy intervals $I_k$ and represent each interval using a
single bath site. $\epsilon_{ik}$ is then given by the center of this interval, while $\tilde V_{ik}^2$ is the area of $\left\{ \bar\mybfDelta_{\text{AI}} (\omega) \right\}_{ii}$ in the given interval.}}
Above, $\bar\mybfG^0_{\text{NI}}(\omega)$ is the NI part of the Green's function defined in Eq.~\ref{eq:G} but already in the transformed space.

\subsection{Impurity solver\label{sec:impuritysolver}}
\label{sec:solver}

To obtain the self-energy of the AIM described above, we solve the impurity model using the recently developed fork tensor product state (FTPS) solver~\cite{Bauernfeind_FTPS_2017,Bauernfeind_FTPSSMO_2018,Bauernfeind_THESIS_2018}. An FTPS is a tensor network based on matrix product states (MPS)~\cite{Schollwoeck_DMRGMPS_2011}, especially suited for impurity models. As FTPS is a Hamiltonian based method, it can only employ a finite but large number of bath states and hence Eq.~\ref{eq:bathParameters} can only be satisfied approximately. 

To solve an AIM with FTPS, we first calculate the ground state $\ket{\psi_0}$ using the density matrix renormalization group (DMRG)~\cite{White_DRMG_1992}. Then by real-time evolution we obtain the retarded Green's function as
\begin{align}\label{eq:FTPS_GF}
    G^{\text{FTPS}}_{ij}(t) &= -i\Theta(t) \bra{ \psi_0 } \{ \opa_{i}(t), \opad{j}(0) \} \ket{ \psi_0 }.
\end{align}
Note that this implies that FTPS is a zero-temperature method. A subsequent Fourier transform to energy space $\omega$ gives access to the impurity spectral function
\begin{align}
    A^{\text{FTPS}}_i(\omega) = - \frac{1}{\pi} \Im G^{\text{FTPS}}_{ii}(\omega)
\end{align}
with
\begin{align}\label{eq:FTPS_GF_omega}
    G^{\text{FTPS}}_{ij}(\omega) = \int e^{ i \omega t - \eta t } G^{\text{FTPS}}_{ij}(t)  dt \;.
\end{align}
The artificial broadening $\eta>0$ is necessary, to avoid finite size effects. In $\omega$ space such a broadening corresponds to a convolution with a Lorentzian of width $\eta$. Although this can have similar effects as a finite temperature, we emphasise that FTPS is a zero temperature method to calculate the $T=0$ spectrum with broadened peaks.

We perform the calculation using the following parameters. Our FTPS tensor network consists of 309 bath sites for each orbital. \rev{Note that we perform the calculations using the Hamiltonian of the AIM in the form given by Eq.~\ref{eq:H_AIM}, i.e., we do not transform onto a nearest neighbor tight binding Wilson chain~\cite{Bulla_NRG_2008,Wolf}.} The truncation at each singular value decomposition (SVD) was $10^{-11}$ during DMRG and $5\cdot 10^{-9}$ during time evolution, where we additionally restrict the maximal tensor index dimensions to $1500$. We choose a Suzuki-Trotter time step  $\Delta t = 0.5$eV to be able to resolve the low-energy part of the spectrum better. This might seem very large, but remember that the energy scales of the Hamiltonian in general are very small ($U=0.5$~eV see below), allowing for a larger time step. Additionally we checked that the result is converged in $\Delta t$. We performed the time evolution up to times $t=800$~eV$^{-1}$ and used a broadening of $\eta=0.005$~eV during Fourier transform (see Eq.~\ref{eq:FTPS_GF_omega}). Furthermore, we made sure that the spectral function of the FTPS solver is consistent with the CTQMC \cite{Werner_CTHYB_2006,Seth_CTHYB_2016} result. We refrain from using CTQMC to solve the impurity model, because it was difficult to reliably discern the splitting of the Kondo resonance from artifacts of the analytic continuation done using the maximum entropy
method \cite{Jarrell_MaxEnt_1996} with an alternative evidence approximation \cite{Linden_MaxEnt_1999} and the preblur formalism \cite{Skilling_MaxEnt_1991}.
Solving the AIM by the FTPS solver leads to the corresponding  Green's function $G^{\text{FTPS}}_{\text{AI}}(\omega)$ of the AIM with approximated bath. Since the number of bath states is large (309 for each orbital, see above), the Green's function with a finite number of bath sites is a very good approximation to the true Green's function of the AIM with the \rev{hybridization} $\bar\mybfDelta_{\text{AI}}(\omega)$. Therefore, we can use the Dyson equation to obtain the self energy of the true AIM:
\begin{align}
\left( G^{\text{FTPS}}_{\text{AI}} \right) ^{-1} (\omega) =
 \left( \bar\mybfG_{\text{AI}}^{0}\right) ^{-1} (\omega)
 -\bar\mybfSigma _{\text{AI}}(\omega) ,
\end{align}
with
\begin{align}
    \bar\mybfG^{0}_{\text{AI}}(\omega) = \frac{1}{\omega + \im 0^+ - \epsilon_{\text{AI,D}} -\bar\mybfDelta_{\text{AI}}(\omega)} \;.
\end{align}

\subsection{Transmission and differential conductance}
\label{sec:trans}

In this section we discuss how to calculate the coherent and an incoherent part of the equilibrium transmission.
Coherent transport is well described by the Fisher-Lee formula,
\begin{align}
 I_\text{coh} = &\frac{2e}{h}\int d\omega (f_\text{L}(\omega)-f_\text{R}(\omega)) \notag \\
 &\times \underbrace{
 \Tr\left[ \Gamma^\text{L}(\omega)G(\omega)\Gamma^\text{R}(\omega)
 \DAG{G }(\omega)\right]
 }_{
 T_{\text{coh}}(\omega)
 } \;.\label{eq:FisherLee}
\end{align}
The factor of 2 accounts for spin and $\Gamma$ denotes the antihermitian part of the corresponding \rev{hybridization} function
\begin{align}
    \Delta^{i}(\omega) &= R^i(\omega) - \frac{\im}{2}\Gamma^i(\omega)
\end{align}
with $i\in\{L,R\}$.
First, let us discuss the separation of the coherent transmission $T_{\text{coh}}$ into three parts: (1) the transmission $T_{\text{NI}}$ of the NI region, (2) the coherent transmission $T_{\text{AI}}$ of the AI, and (3) an interference term $T_{\text{I}}$ between these two.
Obviously, the transmission formula also holds in the transformed space obtained by the transformation matrix $\mybfW$ (see Eq.~\ref{eq:TransformationW}) and $T_\text{coh}$ simplifies to:
\begin{align}
 T_{\text{coh}}(\omega) &= \Tr\left[ 
 \bar\Gamma^\text{L}(\omega)
 \bar G (\omega) 
 \bar\Gamma^\text{R}(\omega)
 \DAG{\bar G }(\omega)
 \right] \label{eq:I:coh:parts} \\
 &= \Tr\left[ \bar\Gamma^\text{L}_{\text{ER}}(\omega)\bar G _{\text{ER}}(\omega)\bar\Gamma^\text{R}_{\text{ER}}(\omega)
 \DAG{\bar G _{\text{ER}}}(\omega)\right] \notag \\
 &= \Tr\left[ \bar\Gamma^\text{L}_{\text{NI}}(\omega)\bar G _{\text{NI}}(\omega)\bar\Gamma^\text{R}_{\text{NI}}(\omega)
 \DAG{\bar G _{\text{NI}}}(\omega)\right] \notag.
\end{align}
The second line holds, since the block matrices $H_{\alpha\beta}$ and $S_{\alpha\beta}$ are negligible, as $\alpha$ and $\beta$ are spatially separated, because the ER includes tip, molecule, and surface of the STM configuration.
We can obtain the third line, because the block matrices $\bar\mybfH_{x,\text{AI}}$ and $\bar\mybfS_{x,\text{AI}}$ with $x\in\{\alpha,\beta\}$ in Eqs.~\ref{eq:barS} and \ref{eq:barH} are zero by construction, see $W_2$ in Eq.~\ref{eq:W2}.
$\bar\Gamma^x_\text{NI}$ with $x\in\{\text{L},\text{R}\}$ are obtained from their respective \rev{hybridization} functions given by 
\begin{widetext}
\begin{align}
 \bar\Delta^\text{L}_{\text{NI}}(\omega) &= ((\omega+\im 0^+) \bar S_{\text{NI},\alpha} - \bar H_{\text{NI},\alpha}) ((\omega+\im 0^+) \bar S_{\alpha\alpha} - \bar H_{\alpha\alpha} - \bar\Delta^\text{L}_{\alpha\alpha}(\omega))^{-1} ((\omega+\im 0^+) \bar S_{\alpha,\text{NI}} - \bar H_{\alpha,\text{NI}}) \notag \\
 \bar\Delta^\text{R}_{\text{NI}}(\omega) &= ((\omega+\im 0^+) \bar S_{\text{NI},\beta} - \bar H_{\text{NI},\beta}) ((\omega+\im 0^+) \bar S_{\beta\beta} - \bar H_{\beta\beta} - \bar\Delta^\text{R}_{\beta\beta}(\omega))^{-1} ((\omega+\im 0^+) \bar S_{\beta,\text{NI}} - \bar H_{\beta,\text{NI}}) \;.
\end{align}
\end{widetext}
The Green's function
\begin{align}
 \bar G _{\text{NI}}(\omega) =& \bar g_{\text{NI}}(\omega) \notag \\
 &+ \underbrace{\bar g_{\text{NI}}(\omega) \bar H_{\text{NI,AI}} \bar G _{\text{AI}}(\omega) \bar H_{\text{AI,NI}} \bar g_{\text{NI}}(\omega)}_{\Delta\bar G _{\text{NI}}} \;
\end{align}
consists of two parts, the noninteracting Green's function $g_{\text{NI}}(\omega)$ and \rev{hybridization} with the interacting Green's function $\bar G _{\text{AI}}(\omega)$.
Inserting $\bar G _{\text{NI}}(\omega)$ into Eq.~\ref{eq:I:coh:parts} gives
\begin{align}
 T_{\text{coh}}(\omega) &= \underbrace{\Tr\left[ \bar\Gamma^\text{L}_{\text{NI}}(\omega)\bar g_{\text{NI}}(\omega)\bar\Gamma^\text{R}_{\text{NI}}(\omega)\bar g_{\text{NI}}^\dagger(\omega)\right]}_{T_{\text{NI}}} \notag \\
 & ~~~+\underbrace{\Tr\left[ \bar\Gamma^\text{L}_{\text{NI}}(\omega)\Delta \bar G _{\text{NI}}(\omega)\bar\Gamma^\text{L}_{\text{NI}}(\omega)(\Delta\bar G _{\text{NI}})^\dagger(\omega)\right]}_{T_{\text{AI}}} \notag \\
 & ~~~+\Tr\left[ \bar\Gamma^\text{L}_{\text{NI}}(\omega)(\Delta \bar G _{\text{NI}})^\dagger(\omega)\bar\Gamma^\text{R}_{\text{NI}}(\omega)\bar g_{\text{NI}}(\omega)\right] \notag \\
 & ~~~+\Tr\left[ \bar\Gamma^\text{L}_{\text{NI}}(\omega)\bar g_{\text{NI}}^\dagger(\omega)\bar\Gamma^\text{R}_\text{{N}I}(\omega)\Delta \bar G _{\text{NI}}(\omega)\right] \notag \\
 &= T_{\text{NI}}(\omega) + T_{\text{AI}}(\omega) + T_{\text{I}}(\omega) \label{eq:coherentTransport_3Parts}\;.
\end{align}
As claimed above, Eq.~\ref{eq:coherentTransport_3Parts} separates the coherent transmission into the three parts $T_{\text{NI}}(\omega)$, $T_{\text{AI}}(\omega)$, and $T_{\text{I}}(\omega)$. Additionally, we can rewrite the coherent transmission over the AI as
\begin{align}
T_{\text{AI}}(\omega) = {\Tr\left[ \bar\Gamma^\text{L}_{\text{AI}}(\omega)\bar G _{\text{AI}}(\omega)\bar\Gamma^\text{R}_{\text{AI}}(\omega)\bar G _{\text{AI}}^\dagger(\omega)\right]}
\end{align}
with
\begin{align}
 \bar\Gamma^\text{L}_{\text{AI}}(\omega) &= \bar H_{\text{AI,NI}} \bar g_{\text{NI}}(\omega) \bar\Gamma^\text{L}_{\text{NI}}(\omega)\bar g_{\text{NI}}^\dagger(\omega) \bar H_{\text{NI,AI}} \notag \\
 \bar\Gamma^\text{R}_{\text{AI}}(\omega) &= \bar H_{\text{AI,NI}} \bar g_{\text{NI}}^\dagger(\omega) \bar\Gamma^\text{R}_{\text{NI}}(\omega)\bar g_{\text{NI}}(\omega) \bar H_{\text{NI,AI}} \;.
\end{align}
The similarity to the Fisher-Lee formula (Eq.~\ref{eq:FisherLee}) indicates that although $T_{\text{AI}}(\omega)$ includes some of the many-body effects, it still treats the many-body system as a noninteracting model with modified (i.e., interacting) Green's function $\bar G_{\text{AI}}(\omega)$. 
The full many-body character of the AIM also gives an incoherent contribution to the current.
For the sake of readability, we omit bars and the index AI in the following expressions but reintroduce them in the final result.
Ness {\it et al.} \cite{Ness_MeirWingreenSimplification_2010} discuss the applicability of the Fisher-Lee formula for a nonequilibrium current
in the presence of interactions and suggest approximating the current flowing from the left lead into the system as
\begin{align}
 I_\text{L} =& \frac{2e}{h}\int d\omega (f_\text{L}(\omega)-f_\text{R}(\omega)) \notag \\
 &\times \Tr\left[ \Gamma^\text{L}(\omega)G(\omega)\Upsilon^\text{R}(\omega)G^\dagger(\omega)\right]
 \label{eq:Iinc}
\end{align}
with $\Upsilon^\text{R}(\omega) = \Gamma^\text{R}(\omega)\Lambda(\omega)$ and
\begin{align}
 \Lambda(\omega)=1 + \Gamma^\text{R}(\omega)^{-1} \frac{f_\text{L}(\omega)-F(\omega)}{f_\text{L}(\omega)-f_\text{R}(\omega)}\Gamma^\text{ee}(\omega)\;.
\label{eq:sepMW4}
 \end{align}
$F(\omega)$ is the nonequilibrium occupation matrix and $\Gamma^\text{ee}(\omega)$ the antihermitian part of the electron self-energy $\Sigma(\omega)$.
The formula shows that interactions not only affect the Green's functions but also renormalize the coupling to the contact through $\Upsilon^\text{R}(\omega)$.
{So far, we only dealt with retarded (r) quantities. For the following derivation we also need the advanced (a) and introduce lesser ($<$) and greater ($>$) quantities} \cite{Ryndyk_Green_2009}.
Ferretti {\it et al.} \cite{Ferretti_MeirWingreenSimplification_2005,Ferretti_abinitiocorrtrans_2005}, Ng {\it et al.} \cite{Ng_lesser_1996}, and Sergueev {\it et al.} \cite{Sergueev_lesser_2002} propose approximating $\Lambda$ using the ansatz
\begin{align}
\Sigma^{>}_{\Delta}(\omega) &= \Delta^{>}(\omega)\Lambda(\omega) \notag \\
\Sigma^{<}_{\Delta}(\omega) &= \Delta^{<}(\omega)\Lambda(\omega) \;.
\end{align}
{$\Delta$ is the total \rev{hybridization}, therefore the tunneling self-energy of the left and the right lead, and $\Sigma_{\Delta}(\omega)$ denotes the self-energy including tunneling and interaction,}
\begin{align}
   \Delta^i(\omega) &= \Delta^{i,\text{L}}(\omega) + \Delta^{i,\text{R}}(\omega) \notag \\
   \Sigma^i_{\Delta}(\omega) &= \Delta^{i,\text{L}}(\omega) + \Delta^{i,\text{R}}(\omega) + \Sigma^i(\omega)\;,
   \label{eq:32}
\end{align}
with $i\in\{\text{r},\text{a},<,>\}$.
Analogous to \cite{Ferretti_abinitiocorrtrans_2005} we define the retarded and advanced self-energies and \rev{hybridization}s, $X\in\{\Sigma,\Delta\}$, as
\begin{align}
 X^{r}(\omega) &:= R(\omega) - \frac{\im}{2}\Gamma(\omega) - \im\delta^+ \notag\\
 X^{a}(\omega) &:= R(\omega) + \frac{\im}{2}\Gamma(\omega) + \im\delta^+  \;,
 \label{eq:selfenergy}
\end{align}
consisting of an hermitian and an anti-hermitian part, $R$ and $\im \Gamma$, respectively.
We added the $\im\delta^+$ term to regularize the inverse of the \rev{hybridization} when $\Gamma(\omega)$ vanishes.
For the various self-energies and \rev{hybridization}s defined so far, the relation $X^>(\omega)-X^<(\omega)=X^r(\omega)-X^a(\omega)$ always holds and by subtracting the two lines in Eq.~\ref{eq:32}, we can write
\begin{align}
 \Lambda(\omega) =& \left( \Delta^r(\omega)-\Delta^a(\omega)\right)^{-1}\left( \Sigma_\Delta^r(\omega)-\Sigma_\Delta^a(\omega)\right) \notag \\ 
                 =& \left( \Gamma^\text{R}(\omega) + \Gamma^\text{L}(\omega) + 2\delta^+\right)^{-1} \notag \\
                 &\times\left( \Gamma^\text{R}(\omega) + \Gamma^\text{L}(\omega) + \Gamma^\text{ee}(\omega) + 2\delta^+\right) \notag \\
                 =& 1 + \left( \Gamma^\text{R}(\omega) + \Gamma^\text{L}(\omega) + 2\delta^+\right)^{-1}\Gamma^\text{ee}(\omega) \;.
\end{align}
We see that $\Lambda$ differs from one only for weak coupling to the leads ($|\Gamma^\text{L}(\omega) + \Gamma^\text{R}(\omega)|\lesssim|\Gamma^\text{ee}(\omega)|$).
As shown in Ref. \cite{Ferretti_abinitiocorrtrans_2005}, $\Lambda(\omega)$ obtained from this this ansatz is exact for {\it nonequilibrium mean field theory} and for the {\it equilibrium many-body} case.

Thus, we can finally write the incoherent part of the transmission
\begin{widetext}
\begin{align}
 I_{\text{L},\text{inc}} =  \frac{2e}{h}\int d\omega (f_\text{L}(\omega)-f_\text{R}(\omega)) \underbrace{\Tr\left[ \bar\Gamma^\text{L}_{\text{AI}}(\omega)\bar G _{\text{AI}}(\omega) \bar\Gamma^\text{R}_\text{AI}(\omega) \left( \bar\Gamma^\text{L}_\text{AI}(\omega) + \bar\Gamma^\text{R}_\text{AI}(\omega) + 2\delta^+\right) ^{-1} \bar\Gamma ^{\text{ee}}_{\text{AI}}(\omega)\bar G _{\text{AI}}^\dagger(\omega) \right]}_{T_{\text{L},\text{inc}}}\;.
\end{align}
\end{widetext}
The total current $I=I_{\text{coh}} + I_{\text{L},\text{inc}}$ and therefore, the total transmission is given by $T(\omega)=T_{\text{coh}}(\omega)+T_{\text{L},\text{inc}}(\omega)$.

For our purposes, the only remaining task is to relate the transmission $T(\omega)$ to the differential conductance, measured by the STM. {As the STM tip usually couples weakly to the molecule, it is reasonable that the voltage $u$ only affects the Fermi function of the right lead describing the STM tip, via a shift of the energy axis. At small temperatures and close to equilibrium ($T(\omega)$ is independent of $u$) we find:}
\begin{align}
 \frac{d}{du} I(u) &\approx \frac{d}{du}\frac{2e}{h}\int_{-\infty}^{\infty} d\omega \; (f_\text{L}(\omega)-f_\text{R}(\omega-u)) \; T(\omega) \notag \\
 &\approx \frac{d}{du}\frac{2e}{h}\int_0^u d\omega \; T(\omega) \propto T(u) \;,
\end{align}
i.e., for small temperatures and voltages, the differential conductance is proportional to the transmission itself.  

\section{Results for CuPC on Au(111) and Ag(111)}
\label{sec:results}

In this section, we use the scheme described above to perform an {\it ab initio} calculation for the electronic transport properties of CuPc on Au(111) and Ag(111), respectively.
In Sec. \ref{sec:DOS} we present the DFT results and combine them with experimental evidence and other theoretical studies to obtain the interaction parameters used for the AIM.
After that, in Sec. \ref{sec:Kondo}, we present the solution of the AIM and estimate the Kondo temperatures of these systems. The different contributions to the transmission are then calculated in Sec. \ref{sec:Transmission}.

\subsection{Density of states and interaction parameters}
\label{sec:DOS}

In this section we estimate the interaction parameters, using a simplified many-body approach (compared to the exact approach outlined in Sec. \ref{sec:impuritysolver}), namely cluster perturbation theory (CPT).
CPT becomes exact for vanishing interaction strength. It is reliable enough for a rough estimation, but it will not be able to describe the Kondo physics appropriately.

\begin{figure*}
\begin{flushleft}
 \hspace{0.2\columnwidth}(a) \hspace{0.83\columnwidth}(b)
\end{flushleft}
 \begin{center}
 \includegraphics[width=0.9\columnwidth,angle=0]{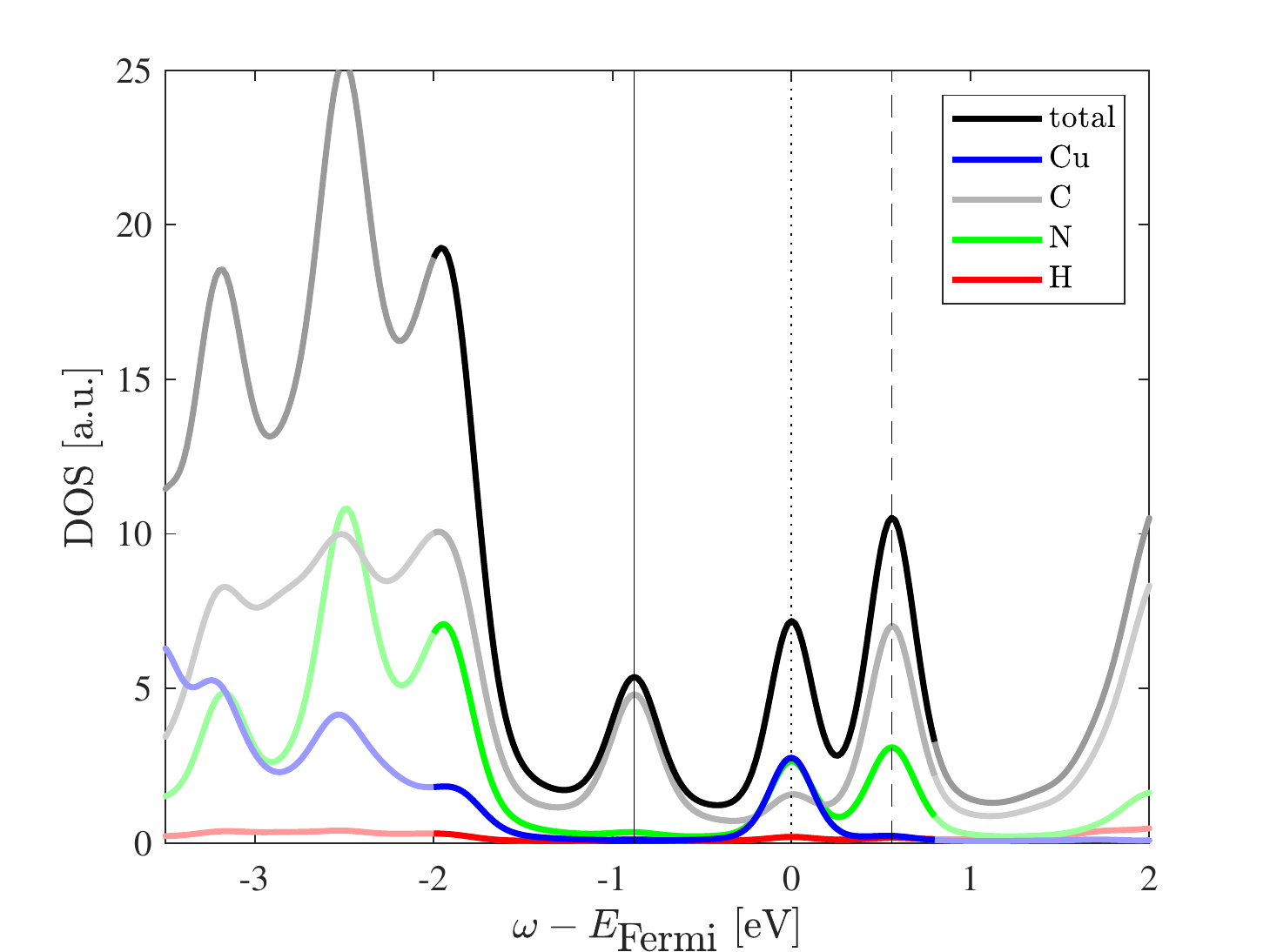} 
 \includegraphics[width=0.9\columnwidth,angle=0]{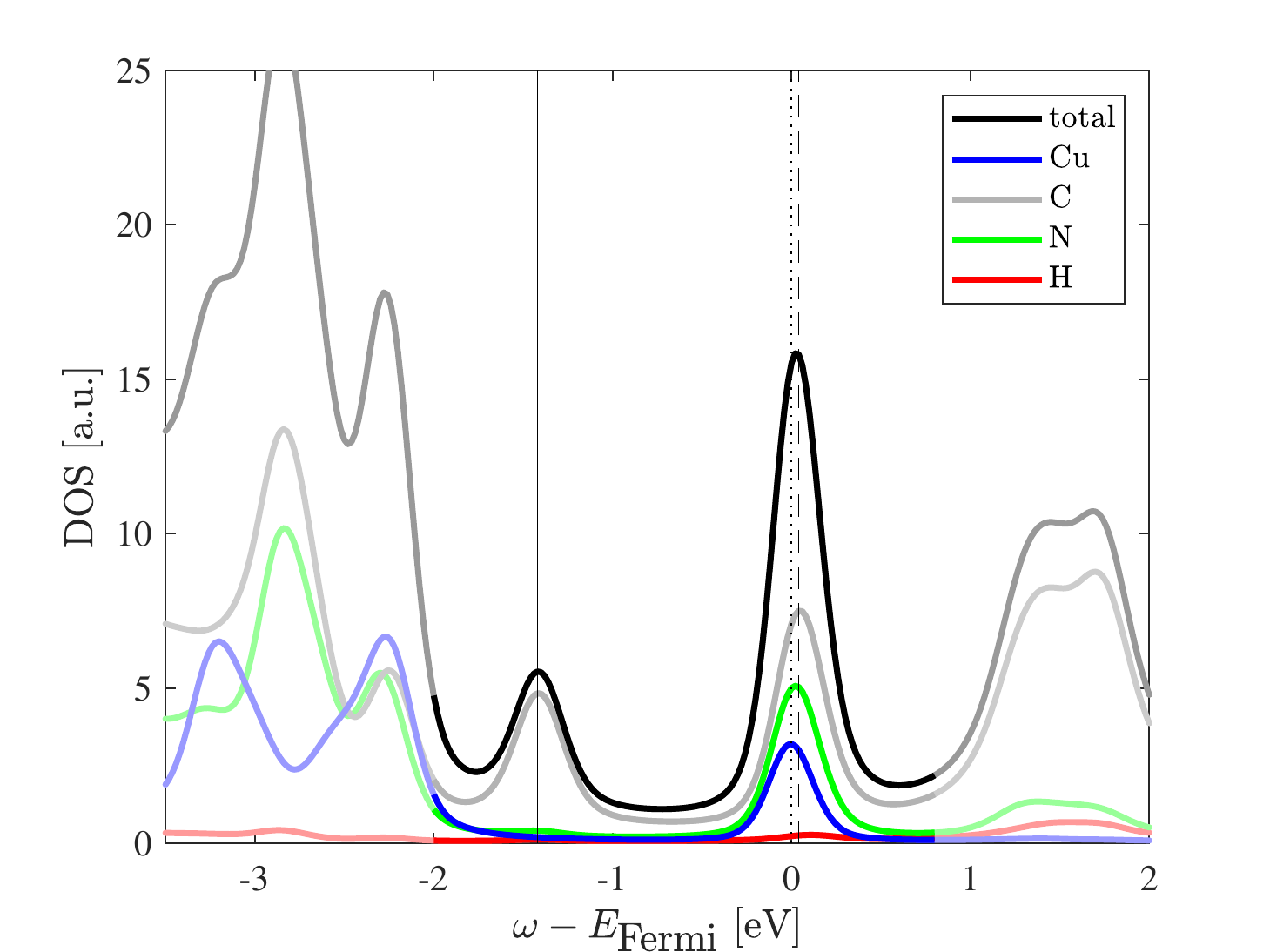} 
 \begin{flushleft}
 \hspace{0.2\columnwidth}(c) \hspace{0.83\columnwidth}(d)
\end{flushleft}
 \includegraphics[width=0.9\columnwidth,angle=0]{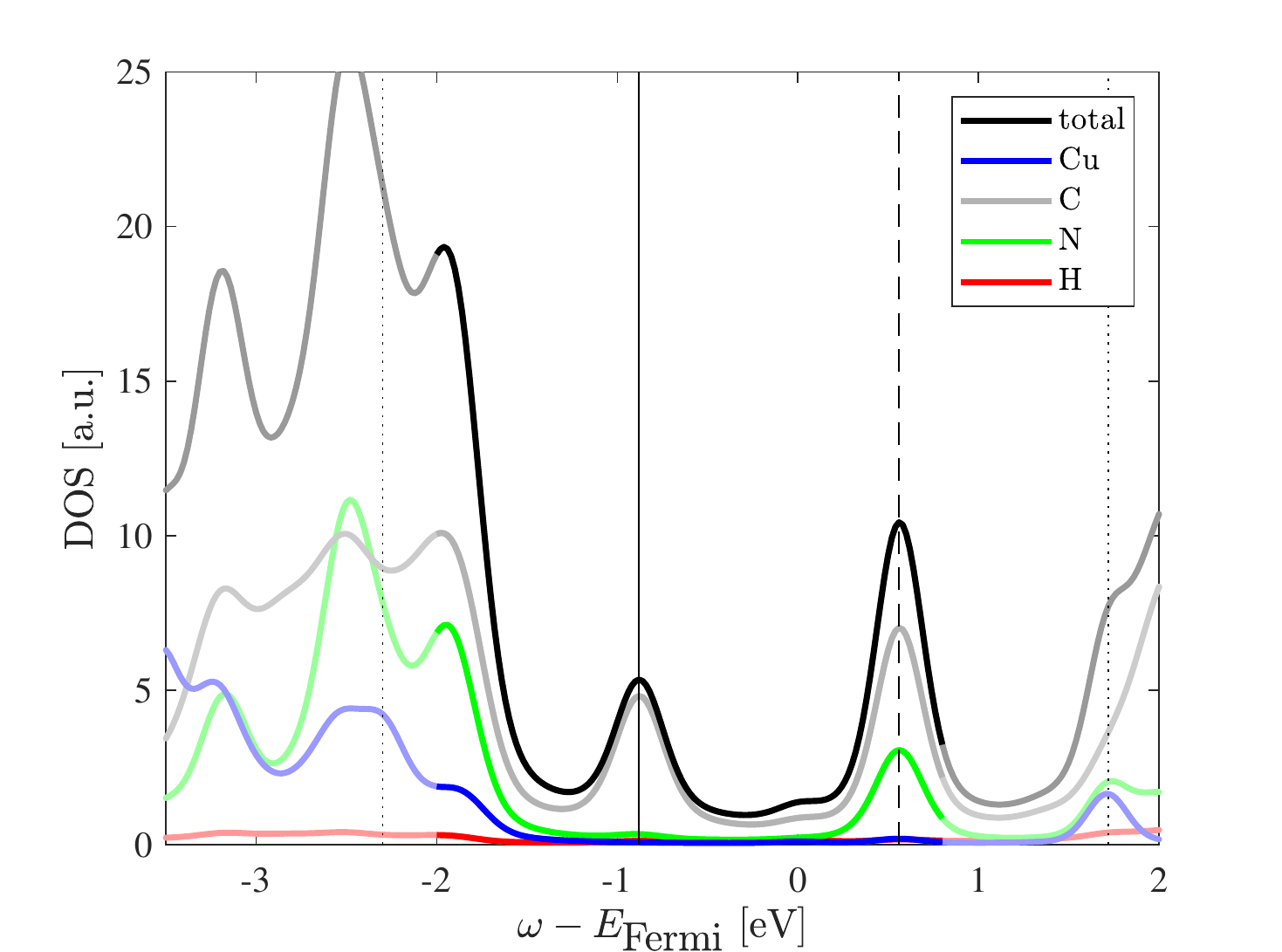} 
 \includegraphics[width=0.9\columnwidth,angle=0]{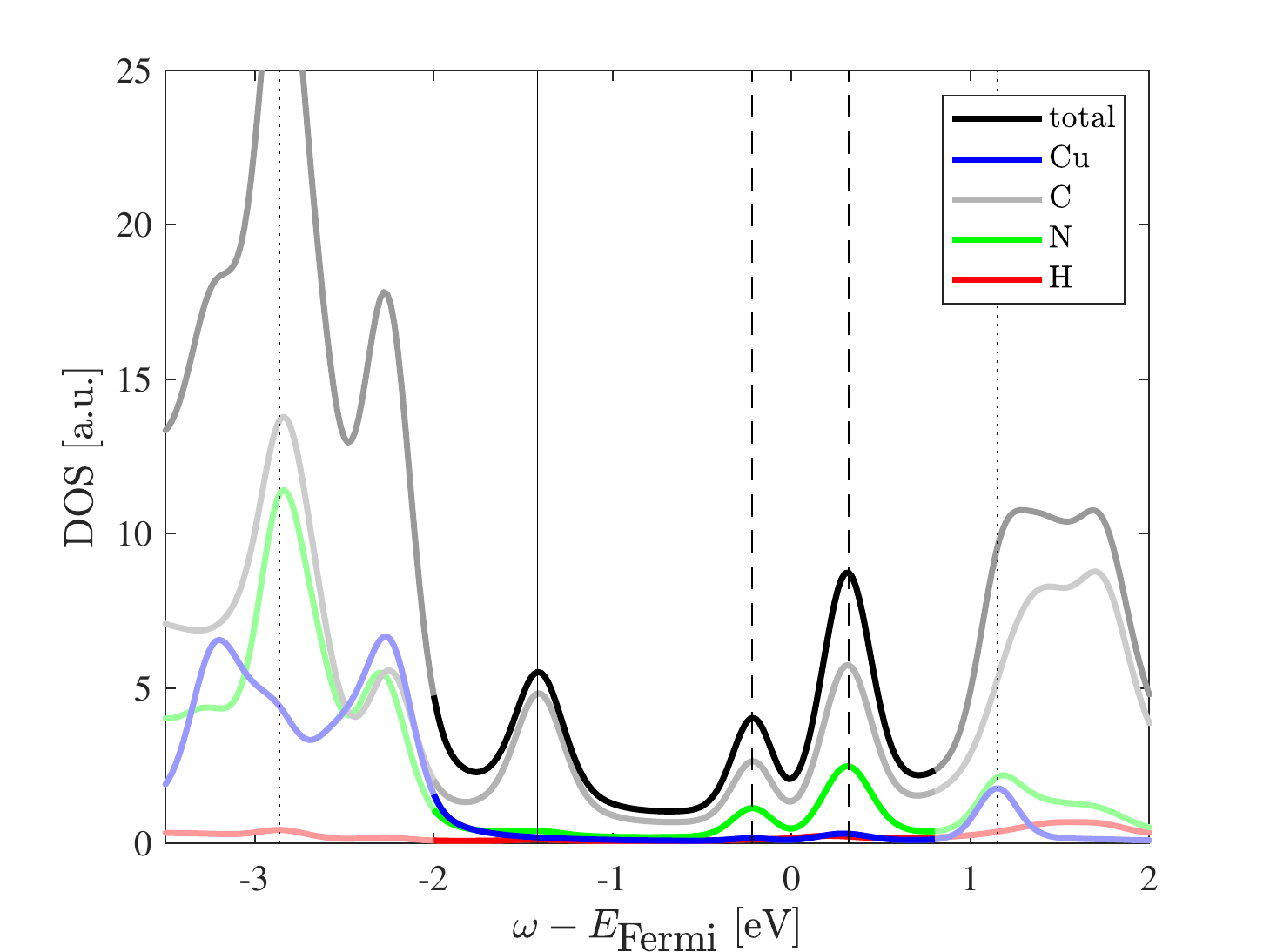} 
 \end{center}
 \caption{Atom resolved DOS of CuPc on Au(111), (a) and (c), and Ag(111), (b) and (d) surface.
 (a) and (b) are the ones obtained in DFT-PBE and (c) and (d) include the interaction term in the CPT approximation.
 We used a $0^+$ of $0.04$ for calculating the DOS and an additional convolution with a Gaussian to obtain a total broadening $\sigma$ of $0.2/\sqrt{2}$.
 \rev{The vertical lines indicate the HOMO (solid line), the position of the $2e_g$ orbitals (dashed lines), and $b_{1g}$ orbitals (dotted lines).}}
\label{fig:CuPc-pDOS}
\end{figure*}

First, we investigate CuPc on Au(111). In Fig. \ref{fig:CuPc-pDOS}(a) we depict the atomic-element resolved DOS obtained from the spin-unpolarized DFT-PBE calculation.
{The orbital directly located at the Fermi energy (partially filled) turns out to have approximately 50 \% copper and 50 \% nitrogen character. The contributions from the carbons and the metal surface are negligible. Therefore, we identify this orbital as the $b_{1g}$ orbital localized in the Cu ion reported in literature, e.g.,} Ref. \cite{Mugarza_CuPc_2012}.
Localization and partial filling ($S=1/2$ for pristine CuPc \cite{Liao_MTPc_2001}) suggest that correlation effects are important for the $b_{1g}$ orbital.
We will model these correlations by adding a Hubbard-type interaction with strength $U_{b_{1g}}$.
To determine its magnitude, we use ultraviolet PES (UPS) spectra obtained in Ref.~\cite{Wruss_Phthalocyanines_2014}. They report the HOMO peak at $-0.81~$eV also seen in our DFT-PBE calculations but at slightly lower energy.
Importantly, the UPS spectra show no additional peak down to $-1.6~$eV, which implies for a Hubbard model at half filling an on-site interaction of $U_{b_{1g}}> 3.2~$eV.
Additionally, DFT calculations using the Heyd-Scuseria-Ernzerhof (HSE) exchange-correlation functional (DFT-HSE) performed in Ref.~\cite{Wruss_Phthalocyanines_2014} suggest $U_{b_{1g}} = 4.0~$eV, which we use in the following.
We use the AMF double counting according to Eq.~\ref{eq:dc}.
As suggested by DFT-HSE calculations the Hubbard satellites are almost symmetric around the Fermi level, which we can achieve using $x=0.85$.
Note that this choice of $x$ does not affect the filling of $n=1$ as suggested by $S=1/2$ of the pristine CuPc.
Since $\bar\Gamma_{b_{1g}} \ll U_{b_{1g}}$, namely $\bar\Gamma_{b_{1g}} = \mathcal{O}(\text{meV})$, using CPT to solve the many-body problem is justified.
The atomic-element resolved DOS thus obtained is shown in Fig. \ref{fig:CuPc-pDOS}(c).
The HOMO peak at around $-0.9~$eV, the spectral weight below $-1.6~$eV, and the absence of the $b_{1g}$ peak at the Fermi level are in good agreement to the DFT-HSE calculations and the UPS spectra of Ref. \cite{Wruss_Phthalocyanines_2014}.

In contrast, for CuPc on Ag(111) photoemission spectroscopy \cite{Ziroff_Phthalocyanines_2012} and for CuPc on Ag(100) scanning tunneling microscopy \cite{Mugarza_CuPc_2011} show a Kondo resonance directly at the Fermi level.
Besides the HOMO peak at $-1.23~$eV, peaks at $-1.74~$eV and $-2.16~$eV and spectral weight below $-2.6~$eV appear in the UPS spectra obtained in Ref.~\cite{Wruss_Phthalocyanines_2014}.
While our DFT-PBE calculation (Fig. \ref{fig:CuPc-pDOS}(b)) shows the HOMO peak at approximately the correct position, other spectral weight can be found already below $-2.0$eV.
Huang et. al \cite{Wruss_Phthalocyanines_2014} demonstrated that this is an artifact of the PBE exchange correlation functional.
As can be seen in figure 6 (b) of Ref.~\cite{Wruss_Phthalocyanines_2014}, the DFT spectral weight below $-2~$eV is shifted down to approximately $-2.6~$eV using HSE instead of PBE.
The authors of  Ref.~\cite{Wruss_Phthalocyanines_2014}
 suggest that the remaining two peaks at $-1.74~$eV and $-2.16~$eV are closely related to the strong interaction between CuPc and Ag(111), especially the feature at $-1.74~$eV.

{In our DFT-PBE calculation for CuPc on Ag(111) three orbitals are located at the Fermi energy and, therefore, partially filled.
As in the case of CuPc on Au(111) we can identify one of them with the $b_{1g}$ orbital.
The remaining two orbitals are nearly degenerated and turn out to consist approximately of 50 \% carbon and 30 \% nitrogen character. Remaining contributions are from the copper ion and the metal surface.
We identify them with the $2e_g$ levels, spatially located mainly at the Pc.}
To model these three correlated orbitals, we choose the AIM Hamiltonian given by Eq.~\ref{eq:modelH} with parameters
\begin{align}
 \mybfU = \begin{pmatrix}
  U_{b_{1g}} & U_{\text{x}} & U_{\text{x}} \\ U_{\text{x}} & U_{2e_g} & U_{2e_g} \\ U_{\text{x}} & U_{2e_g} & U_{2e_g}
 \end{pmatrix} 
\end{align}
and
\begin{align}
 \mybfV = \begin{pmatrix}
  0 & U_{\text{x}}-J & U_{\text{x}}-J \\ U_{\text{x}}-J & 0 & U_{2e_g} \\ U_{\text{x}}-J & U_{2e_g} & 0
 \end{pmatrix}\;.
\end{align}
In analogy to CuPc on Au(111), we take $U_{b_{1g}}=4.0$~eV as the on-site interaction parameter for the $b_{1g}$ orbital.
According to \cite{Korytar_CuPcKondo_2011,Dudarev_CuPcUscreen_1998} the screened interaction $U$ for the $2e_g$ orbitals is between $0.5~$eV and $1.0~$eV on Ag surfaces.
We choose $U_{2e_g}=0.5~$eV, which is also in agreement with the results of  DFT-HSE calculations performed in Ref. \cite{Wruss_Phthalocyanines_2014}.
{In analogy to CuPc on Au(111) we use a factor of $x=0.85$ for the double counting (Eq.~\ref{eq:dc}) in the $b_{1g}$ orbital and $x=1$ for the $2e_g$ system.}
In a first very crude approximation, we neglect correlations between the $b_{1g}$ and the $2e_{g}$ orbitals, therefore $U_\text{x}=J=0$, and solve two independent many-body problems, one for the $b_{1g}$ orbital and the other describing the $2e_g$ orbitals.
In analogy to CuPc on Au(111), we used CPT for the many-body problem of the $b_{1g}$ orbital.
To obtain a first guess for the atomic-element resolved DOS, depicted  in \ref{fig:CuPc-pDOS}(d), we also use the CPT approximation for the many-body problem of the $2e_g$ orbitals. Note that this approximation is not fully justified. Doing so, the DOS including the interaction is qualitatively comparable to the DFT calculations, based on the HSE functional, obtained in Ref. \cite{Wruss_Phthalocyanines_2014}.

\subsection{Kondo temperature and AIM}
\label{sec:Kondo}

Now that all parameters are fixed, we will study the Kondo features and solve the many-body problem accurately by the FTPS solver introduced in Sec. \ref{sec:solver}.
First, let us consider a possible Kondo effect in the $b_{1g}$ orbitals of CuPc on Au(111) and Ag(111).
{For the one-band case in the wide-band limit} \cite{Hewson_Kondo_1997,Haldane_Kondo_1978}, the Kondo temperature is
\begin{align}
 k_\text{B}T_{\text{K,SU(2)}} = \frac{\sqrt{\Gamma U}}{2}\exp{\left( \frac{\pi \epsilon_0 (\epsilon_0 + U)}{\Gamma U} \right) } \;. 
 \label{eq:SU2Kondo}
\end{align}
We already determined the parameters $U_{b_{1g}}=4~$eV and $\epsilon_{b_{1g}}=-2.29~$eV.
In analogy to \ref{eq:selfenergy} the antihermitian part of the \rev{hybridization} relevant for the Kondo effect is given by
\begin{align}
    \bar\Gamma(\omega) = -2\Im \big\{ \bar\Delta_{\text{AI}}(\omega)\big\} \;.
\end{align}
Since $\Gamma$ in Eq.~\ref{eq:SU2Kondo} is in the wide-band limit, and therefore independent of $\omega$, we average ${\bar\Gamma}(\omega)$\footnote{This procedure seems crude, but consider that due to the uncertainty in the DFT part and interaction parameters, we are providing only a rough estimate of $T_\text{K}$.} in the interval $\omega \in [-1,1 ]$, 
\begin{align}
    \Gamma := \frac{1}{2}\int_{-1}^{1}{\bar\Gamma}(\omega)d\omega \;.
\end{align}
{For CuPc on Au(111) $T_{\text{K,SU(2)}} \lesssim 10^{-100}$ with ${{\Gamma}}_{b_{1g}} = 4.7~$meV and, therefore, Kondo features cannot be observed experimentally.
The same is true for CuPc on Ag(111) (${{\Gamma}}_{b_{1g}} = 9.4~$meV, $U_{b_{1g}}=4~$eV, and $\epsilon_{b_{1g}}=-2.85~$eV).
Hence, we do not expect to be able to observer Kondo resonances of the $b_{1g}$ orbital in any of the two systems. Nevertheless, we will show below that the Kondo temperature for the $2e_g$ orbitals in CuPc on Ag(111) is high enough to be visible in experiments. }

\begin{figure}
 \begin{center}
 \includegraphics[width=0.9\columnwidth,angle=0]{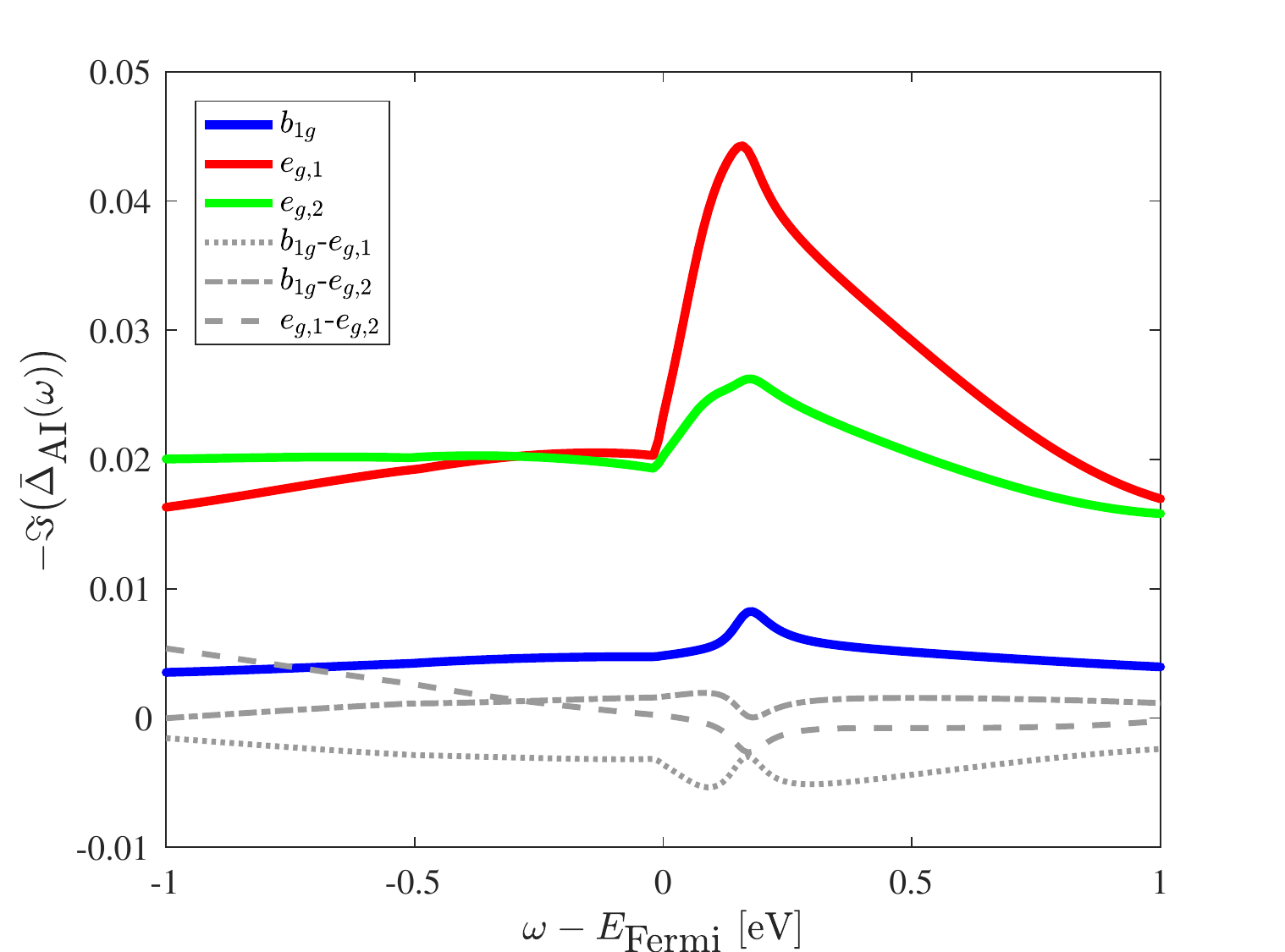} 
 \end{center}
 \caption{Matrix elements of the imaginary part of the \rev{hybridization} $\bar\Delta_{\text{AI}}$ for CuPc on Ag(111).}
\label{fig:hyb}
\end{figure}

Therefore, let us discuss the many-body problem for CuPc on Ag(111) in more depth.
First, we have to determine the missing parameters $J$ and $U_\text{x}$ introduced in Sec. \ref{sec:DOS}(Eq.~\ref{eq:modelH}). These parameters account for the exchange coupling between the $b_{1g}$ and the $2e_g$ electrons and reproduce the Kondo side peaks obtained in Ref. \cite{Mugarza_CuPc_2011}.
According to the energy distance between side peaks and Kondo peak of about $21~$meV we take $J = 25~$meV and $U_{\text{x}} = J$.
The DFT-PBE calculation leads to slightly different on-site energies ($\Delta\epsilon=41~$meV) and \rev{hybridization} functions for the $2e_g$ orbitals, see Fig. \ref{fig:hyb}.
{This difference in the on-site energies $\Delta\epsilon$ causes a similar effect as the exchange coupling $J$, see Eq.~\ref{eq:Hint_egdecoupled} below. 
Therefore, from our {\it ab initio} calculations we cannot conclude whether the Kondo side peaks obtained in} \cite{Mugarza_CuPc_2011} {stem from $\Delta\epsilon$ or $J$.
Hence, we consider only the exchange coupling $J$ and symmetrize the $2e_g$ orbitals ($\Delta\epsilon=0$) and use the same \rev{hybridization} function.}
We also neglect the off-diagonal contributions in the \rev{hybridization} function since they are smaller by a factor of 5 (see Fig. \ref{fig:hyb}) than the diagonal contributions.
Furthermore, because of the strong localization of the $b_{1g}$ orbital, we treat the correlations with the $2e_g$ orbitals in mean field and solve the AIM only in the $2e_g$ subspace using FTPS:
\begin{align}\label{eq:Hint_egdecoupled}
 \opH_{\text{int,}b_{1g}\text{-}e_g} &= J\opn_{b_{1g},\uparrow}\opn_{e_g,\downarrow}+J\opn_{b_{1g},\downarrow}\opn_{e_g,\uparrow} \notag \\
 &\approx J\underbrace{\langle\opn_{b_{1g},\uparrow}\rangle}_{\approx 0}\opn_{e_g,\downarrow}+J\underbrace{\langle\opn_{b_{1g},\downarrow}\rangle}_{\approx 1}\opn_{e_g,\uparrow} \approx J\opn_{e_g,\uparrow} \;,
 \end{align}
where we set $\langle\opn_{b_{1g},\uparrow}\rangle = 0$ and $\langle\opn_{b_{1g},\downarrow}\rangle = 1$. 
For the bath \rev{hybridization}, we choose an energy window $[-1,1]$, see figure \ref{fig:hyb}, and represent this energy range using 309 bath sites for each orbital and spin. Such a large bath is necessary to be able to resolve the fine details of the splitting of the Kondo resonance.
\begin{figure}
 \begin{center}
 \includegraphics[width=0.9\columnwidth,angle=0]{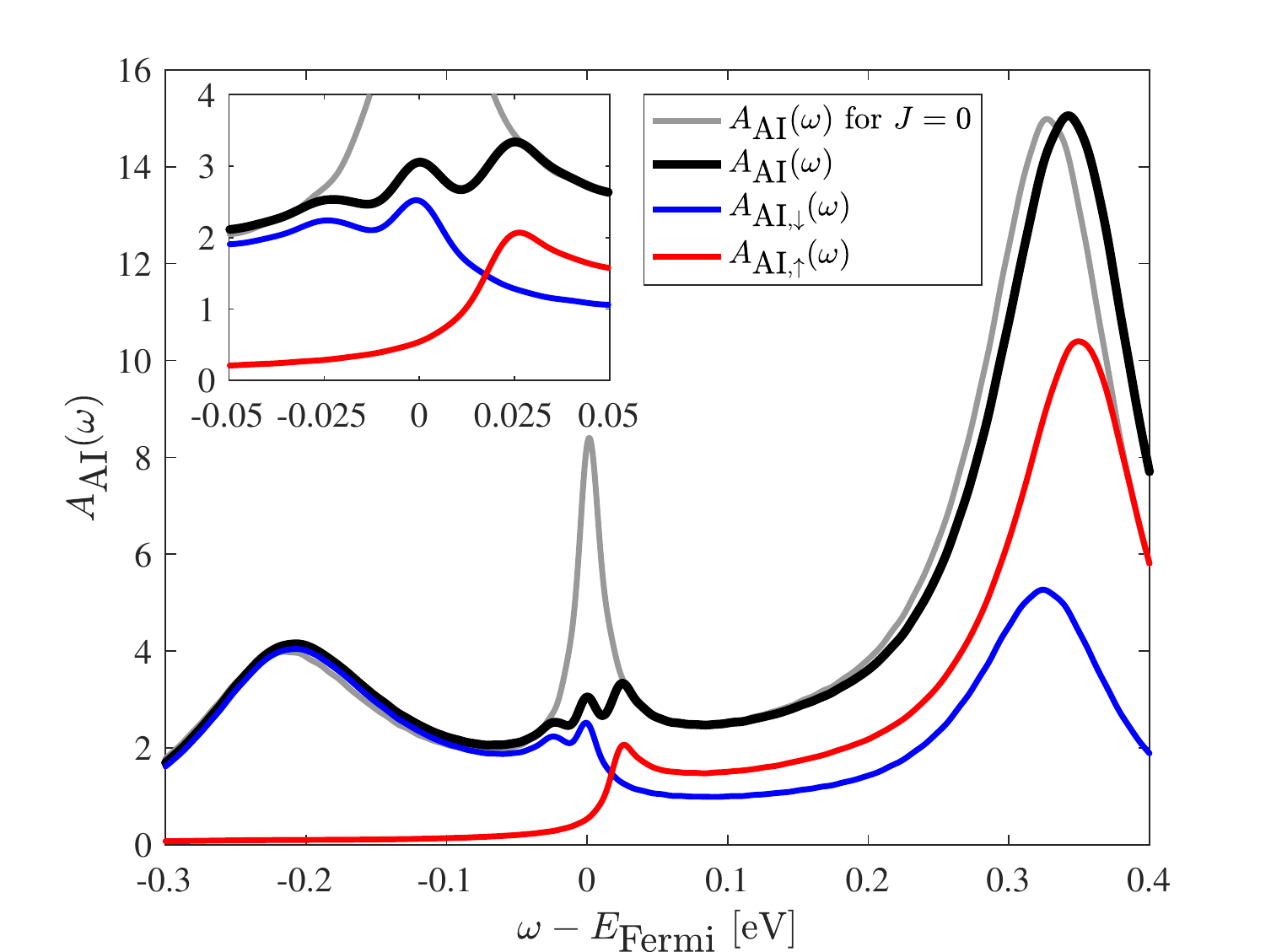}
 \end{center}
 \caption{Spectral function of the Anderson impurity model of CuPc on Ag(111) for $J = 0$ (gray line) and $J = 25~$meV (black line) separated in spin down (blue line) and spin up (red line).}
\label{fig:A_AI}
\end{figure}
The spectral function for the AI orbitals, obtained by FTPS, is shown in figure \ref{fig:A_AI}. {The spectral function for $J=0$ (gray line) shows the familiar scenario consisting of two Hubbard satellites and the Kondo resonance at $0~$eV. An exchange coupling of $J=25~$meV breaks the spin degeneracy by increasing the on-site energy for spin-up electrons according to Eq.} \ref{eq:Hint_egdecoupled} but not the orbital degeneracy.
{Hence, $A_{\text{AI}\downarrow}(\omega)$ (blue line) differs from $A_{\text{AI}\uparrow}(\omega)$ (red line).
Since mainly $A_{\text{AI}\downarrow}(\omega)$ is occupied, the degeneracy of the $2e_g$ orbitals causes an orbital Kondo effect in the spin-down electrons producing the Kondo resonance at $0~$eV. The spin Kondo effect leads to the Kondo satellite peaks at $\omega \approx \pm25~$meV in the total spectral function (black line).} 

Let us discuss the impact of  the symmetry reduction on the Kondo temperature.
In the limit of $J\rightarrow 0$, the spins of the $2e_g$ orbitals are degenerate, causing an SU(4) Kondo effect. For $J\rightarrow \infty$ on the other hand, the two spin-up orbitals are shifted to $+\infty$ and we expect an SU(2) (orbital) Kondo effect from the remaining spin-down degrees of freedom (see Eq.~\ref{eq:Hint_egdecoupled}).
For intermediate values of $J$, we hence expect a situation in between the SU(4) and the SU(2) Kondo regime \cite{Eto_Kondo_2005,Chudnovskiy_SU4KondoZeeman_2005}.
The comparison of the relevant energy scales shows that the exchange coupling $J$ is larger than both Kondo temperatures (SU(2) and SU(4)).
This indicates that the system is closer to the SU(2) than to the SU(4) regime and, therefore, for the Kondo temperature of CuPc on Ag(111) $T_{\text{K,SU(2)}}$ is the better approximation.
The relevant parameters for estimating the Kondo temperature are $U_{2e_g}=0.5~$eV, ${{\Gamma}}_{2e_g} = 44.2~$meV, which is the mean of ${{\Gamma}}_{e_{g,1}}$ and ${{\Gamma}}_{e_{g,2}}$,
and $\epsilon_{2e_g}=-0.20~$eV, being the mean of $\epsilon_{e_{g,1}}$ and $\epsilon_{e_{g,2}}$. 
Equation~\ref{eq:SU2Kondo} {yields $T_{\text{K,SU(2)}}=\{0.02, 1.5, 39\}~$K, where the values are the $\{25, 50, 75\}~$\%-quantile. The quantiles are determined by assuming a Gaussian distribution for $\Gamma$, $U$ and $\epsilon_0$ centered at the value obtained in the previous section and with a standard deviation which is $50~\%$ of the modulus of that value\footnote{This magnitude of the error accounts for uncertainties due to approximations in DFT and the estimation of the interaction parameters.}.}
{We emphasise that the Kondo temperature depends sensitively (exponentially) on the relevant parameters and therefore getting the correct order of magnitude for $T_\text{K}$ is already a remarkable result.}
A closed analytical expression for the Kondo temperature of the SU(4) symmetrical Anderson model is given in Appendix \ref{sec:App:SU4} according to Ref.~\cite{Filippone_SU4Kondo_2014}.
{Application of this formula yields} 
$T_{\text{K,SU(4)}}=\{3, 25, 84\}~$K. Both, $T_{\text{K,SU(2)}}$ and $T_{\text{K,SU(4)}}$ are consistent with the experimentally obtained Kondo temperature {for CuPc on Ag(100)} of $T_K=27~$K \cite{Mugarza_CuPc_2011}.

To be able to obtain values for the Kondo temperature comparable to experiment, Korytár {\it et al.} \cite{Korytar_CuPcKondo_2011} rescaled $\epsilon$ and $\Gamma$. 
The \rev{hybridization} ${{\Gamma}}_{2e_{g}}$ obtained in our calculation is smaller than $\Gamma$ obtained in Ref. \cite{Korytar_CuPcKondo_2011} and therefore gives a better estimate of the Kondo temperature. Hence, we suggest an imprecise adsorption geometry as possible origin of the necessity of this rescaling procedure.
We are using the geometry configuration obtained by Huang {\it et al.} \cite{Wruss_Phthalocyanines_2014}. As shown in our calculation it is possible to get at least the correct order of magnitude for the Kondo temperature from {\it ab initio} calculations.

\subsection{Transport properties}
\label{sec:Transmission}

\begin{figure}
\begin{flushleft}
 \hspace{0.05\columnwidth}(a) \hspace{0.45\columnwidth}(b) 
\end{flushleft}
 \begin{center}
 \includegraphics[width=0.45\columnwidth,angle=0]{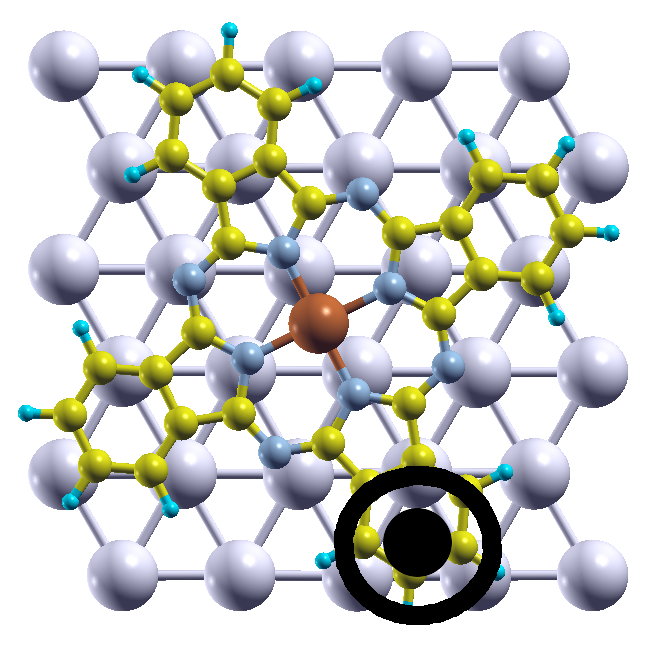}
   \hspace{0.05\columnwidth}
 \includegraphics[width=0.45\columnwidth,angle=0]{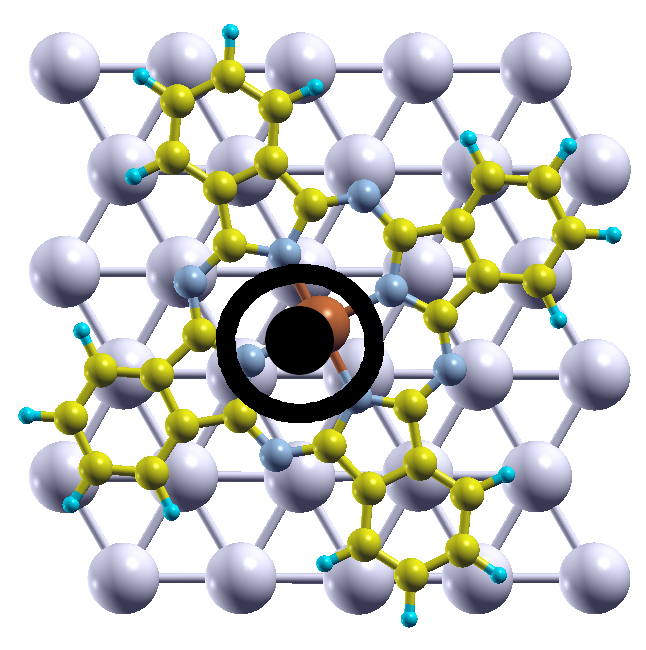}
 \end{center}
 \caption{The tip positions I (a) and II (b) of the STM tip (black sphere) on the molecule on the Ag(111) surface. The pictures are drawn with XCrySDen \cite{Kokalj_xcrysden_2003}.}
\label{fig:tippos}
\end{figure}

As discussed in Sec.~\ref{sec:DOS} {the $2e_g$ orbitals contain 50 \% contribution from the carbon atoms, 30 \% from the nitrogen atoms, while the remaining contributions are from the copper ion and the metal surface.
Whether a Kondo feature can be observed in the differential conductance measurements with an STM therefore depends on the position of the tip. In particular, if the tip is placed above the benzene rings, we expect to observe a Kondo resonance, which should be absent if the tip is above the Cu atom} (see Fig.~\ref{fig:tippos}).
Therefore, let us discuss the transport properties for the two tip positions used in the experiment performed in Ref.~\cite{Mugarza_CuPc_2011,Mugarza_CuPc_2012}(also shown in  Fig.~\ref{fig:tippos}).

\begin{figure}
\begin{center}
 \includegraphics[width=0.9\columnwidth,angle=0]{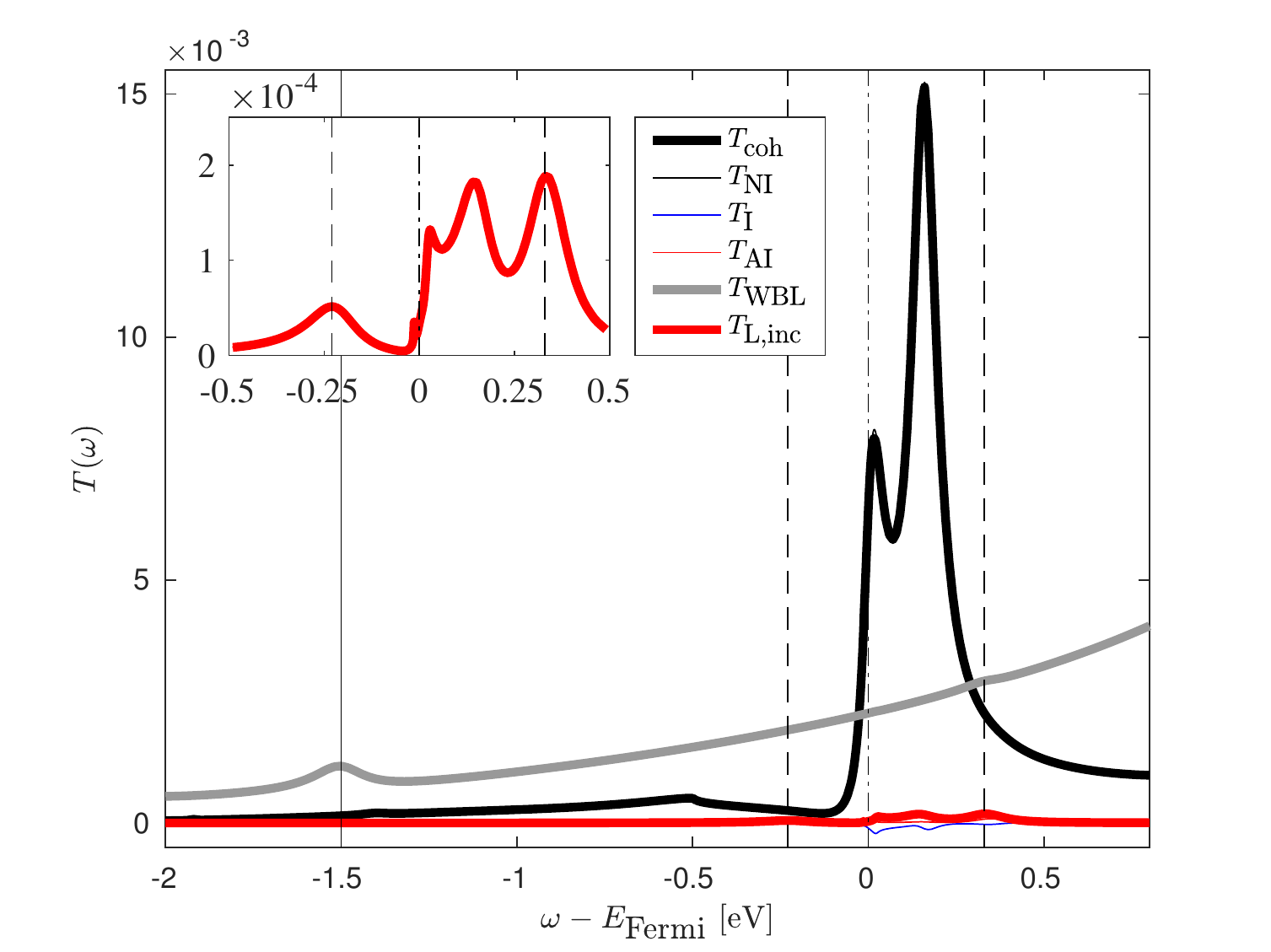}
 \end{center}
 \caption{Transmission of CuPc on Ag(111) in tip position I. The vertical lines mark the position of the HOMO resonance (solid line), the Kondo resonance at $0~$eV (dash-dotted line), and the positions of the Hubbard satellites (dashed lines).}
 \label{fig:CuPc-AgTransmission_I}
\end{figure}

Figures~\ref{fig:CuPc-AgTransmission_I} and \ref{fig:CuPc-AgTransmission_II} show the calculated transmissions for the two STM tip positions. In addition to the total transmission we also show its different contributions, as derived in Sec. \ref{sec:trans}:
The coherent part $T_{\text{coh}}$, consisting of $T_{\text{NI}}$, $T_{\text{AI}}$, and $T_{\text{I}}$,  as well as the incoherent part $T_{\text{L,inc}}$.
A difficulty in the comparison with the published experimental results in Ref.~\cite{Mugarza_CuPc_2011} is that the authors performed a background subtraction for the STM differential conductance measurements, as proposed in Ref.~\cite{Wahl_STMBackground_2008}. 
{The authors introduced the background subtraction} to obtain the transmission of the molecule only, without effects stemming from the tip or the surface.
To take the background subtraction into account in our calculations, we introduce a wide-band limit (WBL) approximation.
{Therefore we define the molecular region ($\text{MR}\subseteq\text{IR}$) consisting of all atomic orbitals located at the CuPc molecule.
Using a modified \rev{hybridization} $\Gamma^{\text{x,WBL}}_{\text{MR}}$ for $x\in\{L,R\}$ leads to the following transmission formula:}
\begin{align}
 T_{\text{WBL}}(\omega) &= \Tr\left[ \Gamma^{\text{L,WBL}}_{\text{MR}} \mybfG^{\text{WBL}}_{\text{MR}}(\omega)\Gamma^{\text{R,WBL}}_{\text{MR}}{ \mybfG^{\text{WBL}}_{\text{MR}}}^\dagger(\omega)\right] \;,
\label{eq:I:WBL}
\end{align}
where direct tunneling from the surface to the tip is neglected.
In the WBL approximation, we replace the imaginary parts of the \rev{hybridization} functions by the constant
\begin{align}
   \Gamma^{x\text{,WBL}}_{\text{MR}} =  \int_{-1/2}^{1/2} \Gamma^x_{\text{MR}}(\omega) d\omega \;.
\end{align}
To obtain the corresponding real parts, we use the Kramers-Kronig relations. The Green's function,
\begin{widetext}
\begin{align}
G^{\text{WBL}}_{\text{MR}}(\omega) &= \left( \omega\mybfS_\text{MR} -\mybfH_\text{MR} - \mybfDelta^{\text{WBL}}_\text{L,MR}(\omega) -  \mybfDelta^{\text{WBL}}_\text{R,MR}(\omega) - \mybfSigma _\text{MR}(\omega)\right) ^{-1} \;,
\end{align}
\end{widetext}
includes only the MR part of the self-energy.

\begin{figure}
 \begin{center}
 \includegraphics[width=0.9\columnwidth,angle=0]{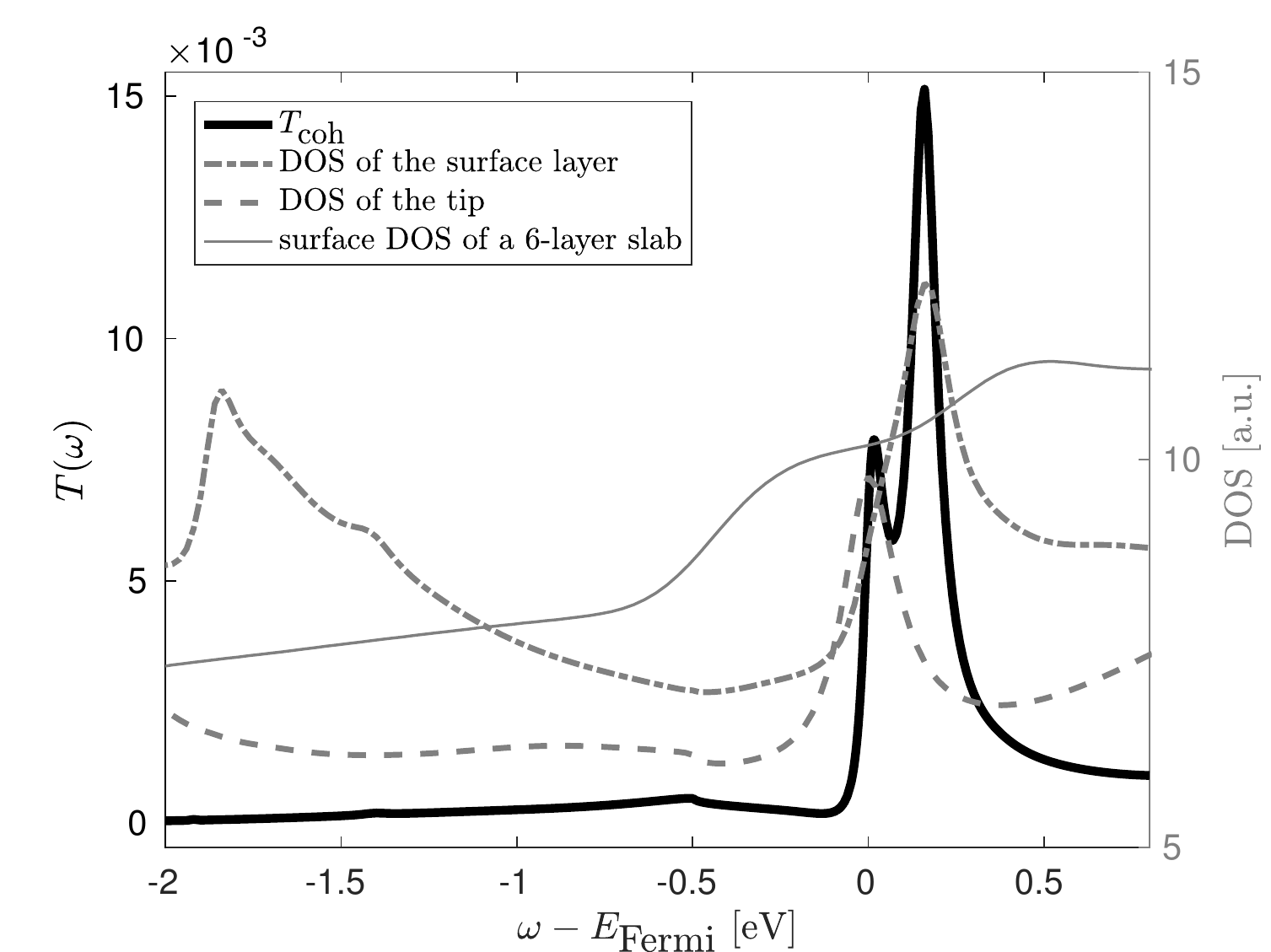}
 \end{center}
 \caption{Coherent transmission of the STM configuration in tip position I (left axis) and the DOS projected on the atomic orbitals of the tip and the surface layer and the surface DOS of a 6-layer slab calculation (right axis).}
\label{fig:pDOS_surf}
\end{figure}

\begin{figure}
\begin{center}
 \includegraphics[width=0.9\columnwidth,angle=0]{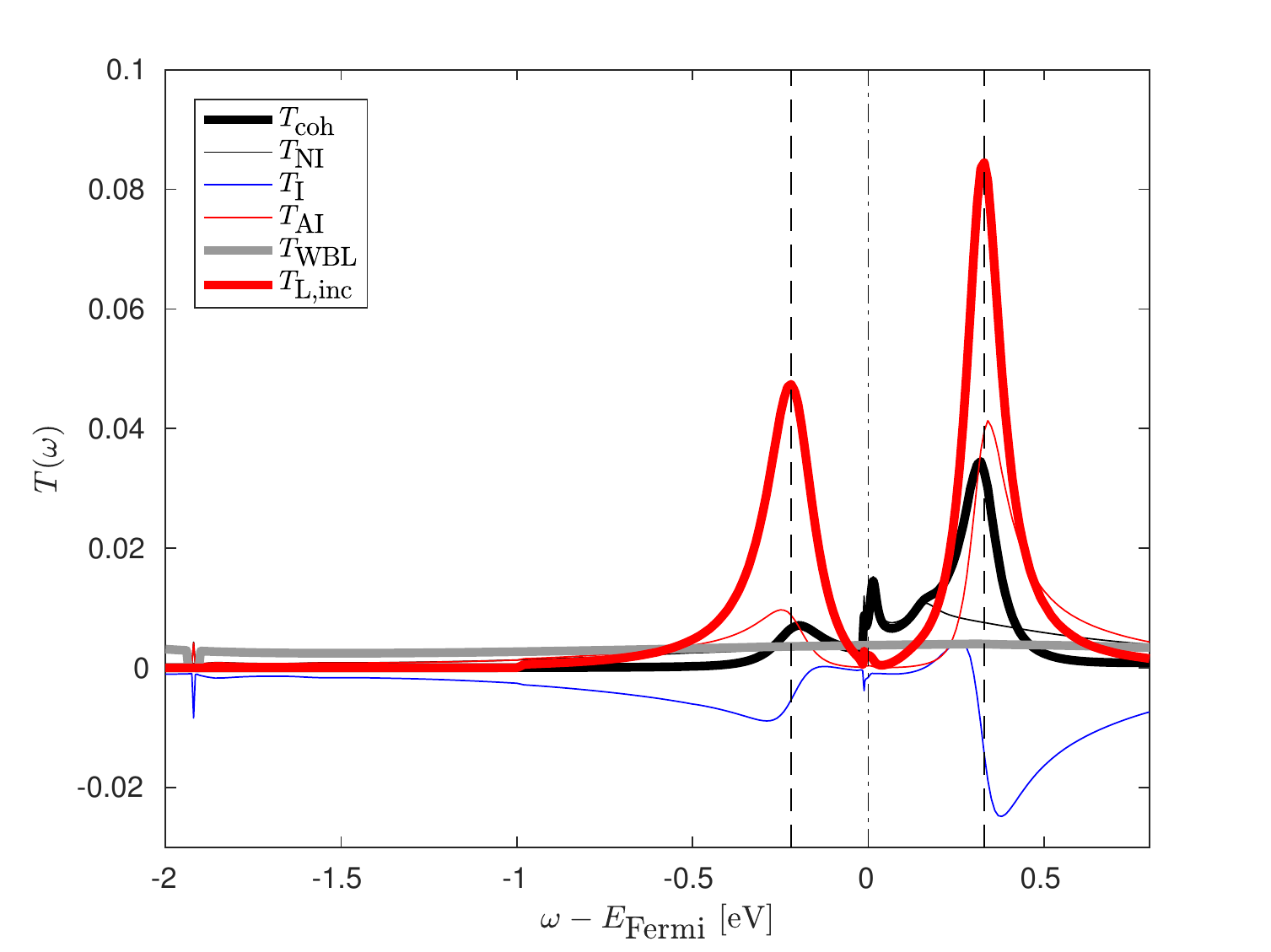}
 \end{center}
 \caption{Transmission of CuPc on Ag(111) in tip position II. The vertical lines mark the Kondo resonance at $0~$eV (dash-dotted line) and the positions of the Hubbard satellites (dashed lines).}
\label{fig:CuPc-AgTransmission_II}
\end{figure}

Figure \ref{fig:CuPc-AgTransmission_I} shows the resulting transmission calculated for tip position I, obtained from the {\it ab initio} calculation, as well as in the WBL. 
First of all, we observe in the {\it ab initio} case that the largest contribution to the coherent transmission $T_\text{coh}$ (black line) is from $T_\text{NI}$ (thin black line, mostly covered by the black line). It has only small contributions from $T_{\text{AI}}$ (thin red line, covered by the red line) and $T_{\text{I}}$ (thin blue line).
Moreover, the coherent transmission $T_\text{coh}$ is dominated by two peaks at energies $0.02~$eV and $0.16~$eV, respectively. 
These peaks are missing in the WBL (gray line) indicating that they cannot be attributed to the pristine molecule.
To underpin this interpretation, we present in Fig. \ref{fig:pDOS_surf}  the projected DOS of the surface (dash-dotted line) and the tip (dashed line).
 {We observe  that the projected DOS of the surface layer and tip show peaks at $0.02~$eV and $0.16~$eV, respectively, coinciding with the peaks in the coherent transmission (black line).
Additional six-layer slab calculations (gray line) of the pristine Ag(111) surface with a vacuum gap of 10~\AA~and an appropriate number of k-points, show that these peaks in the DOS are an artifact of the finite p($6\times 5$) surface.
To avoid these artifacts, we would have to increase the number of atoms in the super cell, which is computationally very demanding and would most likely not provide additional information.}

In the WBL, the transmission $T_\text{WBL}$ in Fig.~\ref{fig:CuPc-AgTransmission_I} is much smoother and clearly shows the HOMO peak, marked by a vertical line at about $-1.5eV$, which is in agreement with the experiment in Refs.~\cite{Mugarza_CuPc_2011,Mugarza_CuPc_2012}. 
The incoherent part of the transmission for position I of the tip is shown in Fig.~\ref{fig:CuPc-AgTransmission_I} as a red line  (see also the inset). Apart from the two peaks induced by the surface and the tip, we find Hubbard satellites (dashed vertical lines) and the Kondo feature (dash-dotted vertical line). The positions of these peaks are also in agreement with the experiment.

Finally, in Fig.~\ref{fig:CuPc-AgTransmission_II}, we present  the results obtained for tip position II. 
We find that the overall transmission is larger by one order of magnitude.
The coherent transmission $T_\text{coh}$ (black line) has contributions from $T_\text{NI}$ (thin black line, mostly covered by the black line), from $T_{\text{AI}}$ (thin red line) and the interference part $T_{\text{I}}$ (thin blue line).
$T_\text{NI}$ consists mainly of the surface and the tip features discussed above and $T_{\text{AI}}$ of the Hubbard satellites, again marked with dashed vertical lines.
There is almost no structure in $T_\text{WBL}$ (gray line).
The incoherent transmission (red line) shows Hubbard satellites, marked with dashed vertical lines, and the Kondo feature, dash-dotted vertical line.
The Kondo resonance is small compared to the height of the Hubbard bands.

There are two big differences between the results for tip positions I and II.
First, the HOMO peak at about $-1.5eV$ appears in tip position I and not in tip position II (compare gray lines in Figs. \ref{fig:CuPc-AgTransmission_I} and \ref{fig:CuPc-AgTransmission_II}).
Second, the Kondo resonance is very pronounced in tip position I, while at tip position II the height of the Hubbard satellites is much larger than the height of the Kondo resonance (compare red lines in Figs. \ref{fig:CuPc-AgTransmission_I} and \ref{fig:CuPc-AgTransmission_II}).
Both findings are in agreement with the experiment {for CuPc on Ag(100) in Ref.} \cite{Mugarza_CuPc_2011}.

\section{Conclusions}
\label{sec:conclusion}

We investigated equilibrium and transport properties of a copper phthalocyanine (CuPc) molecule adsorbed on Au(111) and Ag(111).
Apart from the usual coherent contributions to the transmission, several localized partially filled (strongly-correlated) orbitals also lead to an incoherent part.
%
As the starting point for our {\it ab initio} calculation we used the adsorption geometry obtained by Huang {\it et al.}
\cite{Wruss_Phthalocyanines_2014} and performed DFT calculations that describe the coherent part of the transmission reasonably well.
To tackle the strongly-correlated part, we first used the transformation scheme described by Droghetti {\it et al.} \cite{Droghetti_TOV_2017} to obtain an Anderson impurity model (AIM) based on the DFT calculations. We estimate the interaction parameters from theoretical and experimental data,  Refs.~\cite{Mugarza_CuPc_2011,Wruss_Phthalocyanines_2014}.
For CuPc on Au(111) there is one unpaired spin in the $b_{1g}$ orbital located at the copper ion. Whereas for CuPc on Ag(111) there is an additional unpaired electron in two almost degenerated $2e_g$ orbitals.
In both systems the coupling between the $b_{1g}$ orbital and the remaining system is weak and, therefore, the Kondo temperature for the $b_{1g}$ orbital is very small.
Hence, no Kondo resonance is found in experiments for CuPc on Au(111).

This is different for CuPc on Ag(111), where the AIM consists of three partly filled orbitals with two electrons, the $b_{1g}$ orbital and the $2e_g$ orbitals. While the Kondo temperature of the $b_{1g}$ is still very small, the other electron in the $2e_g$ orbitals shows a measurable Kondo resonance. We solve the corresponding AIM obtained by the transformation scheme using cluster perturbation theory for the $b_{1g}$ subspace and the fork tensor product state solver \cite{Bauernfeind_FTPS_2017} for the $2e_g$ subspace. To combine the two subspaces, we treat the correlations between the subspaces on a mean field level. The mean field coupling reduces the SU(4) symmetry of the $2e_g$ subspace into a SU(2) symmetry of the orbital degrees of freedom. This in turn leads to a Kondo effect with a symmetry somewhere between SU(2) and SU(4) for CuPc on Ag(111).

Since the Kondo temperature depends sensitively on the \rev{hybridization} of the molecule with the surface, the adsorption geometry is very important in the DFT calculation. In our opinion this is the reason why previous {\it ab initio} studies by Korytar et. al~\cite{Korytar_CuPcKondo_2011} were not able to obtain correct estimates of the magnitude of the Kondo temperature without rescaling parameters.

Indeed, using the relaxed geometries yields reliable {\it ab initio} estimates of the Kondo temperature and reproduces the qualitative behavior of the differential conductance found in the STM measurements of Refs. \cite{Mugarza_CuPc_2011,Mugarza_CuPc_2012}. 

\section{Acknowledgments}

The authors like to thank Elisabeth Wruss for introducing us to her DFT results of phthalocyanines on different surfaces, Thomas Taucher for discussions about {\sc TranSIESTA}, and Gernot Kraberger and Robert Triebl for discussions concerning the many-body problem.
{This work was partially supported by the Austrian Science Fund (FWF) under Grant No. P26508, the SFB-ViCoM F4104, and through the START program Y746, as well as by NAWI Graz.}
The DFT calculations were partly done on high performance computing resources of the ZID at TU Graz.

\section{Appendix}

\rev{\subsection{Density functional calculation details}
\label{sec:App:DFT}

The DFT calculations are performed with {\sc SIESTA} \cite{Soler_Siesta_2002} and Tran{\sc SIESTA} \cite{Brandbyge_TranSiesta_2002}.
We use the Perdew-Burke-Ernzerhof (PBE) \cite{Perdew_BPE_1996} functional, which is a generalised gradient approximation (GGA) functional.
To suppress periodicity effects, we perform the calculations at the $\Gamma$ point, except for the electrode calculations, where we use $100$ $k$ points in the transport direction, with one electrode {unit cell} consisting of six metal layers. 
For an appropriate description of the surface, we have to ensure that the super cell is large enough parallel to the surface in order to justify a $\Gamma$ point calculation. For computational reasons we restrict ourselves to the p($6\times5$) surface and discuss possible consequences in Sec.~\ref{sec:Transmission}.
We use {$300$ Ry mesh-cutoff}, and an electronic temperature of $5~$meV.
For the H, C, N, and Cu atoms we use nonrelativistic norm-conserving pseudopotentials \cite{Troullier_TMPseudos_1991} from the Abinit's pseudo database~\footnote{\url{https://departments.icmab.es/leem/siesta/Databases/Pseudopotentials/periodictable-gga-abinit.html}} and for Au and Ag relativistic pseudopotentials as recommended by Rivero {\it et al.}~\cite{Rivero_pseudos_2015}.
For the basis set we restrict ourselves to the standard single-zeta basis plus polarization (SZP) and double-zeta basis plus polarization (DZP) basis sets with an {energy shift of $0.01~$Ry}.
We perform our calculations using an SZP {basis} set for the bulk atoms and a DZP basis set for the atoms in the molecule, the first 2 layers of the metal surface as well as the tip.
Additionally, we use an extended cutoff radius of $7.5~$\AA~for the first zeta basis functions of the four atoms at the tip for calculating the transmissions.
We successfully benchmark the {pseudopotentials} and basis sets calculating the bulk band structure with {\sc SIESTA} and {\sc Quantum Espresso} \cite{QE_2009}.
We apply Tran{\sc SIESTA} in equilibrium and choose the complex contour consisting of a circular part from $-40~$eV to $-10k_BT$ and a tail to infinity.
The imaginary part of the Fermi function tail when crossing the Fermi level is chosen to be {$2.5~$eV} and the complex contour consists of 
Gauss-Legendre quadrature with {96 points is used for the circle} and a Gauss-Fermi quadrature with {16 points is used for the tail}.}

\subsection{Kondo temperature of the SU(4) symmetrical Anderson model}
\label{sec:App:SU4}

For completeness we provide the analytical expression for the Kondo temperature of the SU(4) symmetrical Anderson model derived in \cite{Filippone_SU4Kondo_2014} via a path integral approach,
\begin{align}
 k_\text{B}T_{\text{K,SU(4)}} =& Uf\left( \frac{\epsilon_0}{U}\right) \left( \frac{-2\Gamma U}{\pi \epsilon_0 (\epsilon_0+U)} \right)^{\frac{1}{4}} \notag \\
 &\times \exp{\left( \frac{\pi \epsilon_0 (\epsilon_0 + U)}{2\Gamma U} \right) }
\end{align}
with
\begin{align}
 f(x) = \left( -x(x+1)^3\right)^{\frac{1}{4}}\exp{\left( g(x) \right) }
 \end{align}
and
\begin{align}
 g(x)= \frac{1}{4}\frac{3x-2}{x+2}-\frac{x^2}{2}\frac{(x^2+3x+3)}{(x+2)^2}\ln{\left( \frac{2x+3}{x+1}\right) } \;.
\end{align}

\section*{References}

\bibliographystyle{apsrev4-1}
\bibliography{literatur}

\end{document}